\documentclass[fleqn,usenatbib]{mnras}

\usepackage{newtxtext,newtxmath}
\usepackage[T1]{fontenc}

\DeclareRobustCommand{\VAN}[3]{#2}
\let\VANthebibliography\thebibliography
\def\thebibliography{\DeclareRobustCommand{\VAN}[3]{##3}\VANthebibliography}

\usepackage{natbib}
\usepackage{graphicx}	% Including figure files
\usepackage{amsmath}	% Advanced maths commands

\title[Solar Coronal Heating by Alfv\'{e}n Waves]{Solar Coronal Heating: Role of Kinetic and Inertial Alfv\'{e}n Waves in Heating and Charged Particle Acceleration}

\author[Ayaz et al.]{
Syed Ayaz$^{1}$\thanks{E-mail: sa0173@uah.edu (UAH), syedayaz263@gmail.com},
Gary P. Zank$^{1}$,
Imran A. Khan$^{2,3}$, Yeimy J. Rivera $^{4}$,
Andreas Shalchi $^{5}$, and
L.-L. Zhao $^{1}$
\\
$^{1}$Department of Space Science and CSPAR, University of Alabama in Huntsville, Huntsville, AL 35899, USA\\
$^{2}$Department of Space Sciences, Institute of Space Technology, Islamabad 44000, Pakistan\\
$^{3}$Space and Astrophysics Research Lab (SARL), National Center of GIS and Space Applications (NCGSA), Islamabad 44000, Pakistan\\
$^{4}$ Center for Astrophysics, Harvard and Smithsonian, Cambridge, MA 02138, USA\\
$^{5}$ Department of Physics and Astronomy, University of Manitoba, Winnipeg, Manitoba R3T 2N2, Canada
}

\date{Accepted XXX. Received YYY; in original form ZZZ}

\pubyear{2025}

\begin{document}
\label{firstpage}
\pagerange{\pageref{firstpage}--\pageref{lastpage}}
\maketitle

\begin{abstract}
A comprehensive understanding of solar coronal heating and charged particle acceleration remains one of the most critical challenges in space and astrophysical plasma physics. In this study, we explore the contribution of Alfvén waves—both in their kinetic (KAWs) and inertial (IAWs) regimes—to particle acceleration processes that ultimately lead to coronal heating. Employing a kinetic plasma framework based on the generalized Vlasov-Maxwell model, we analyze the dynamics of these waves with a focus on the perpendicular components (i.e., across the magnetic field lines) of the Poynting flux vectors and the net resonance speed of the particles.

Our findings reveal that both the magnitude and dissipation rate of the Poynting flux vectors for KAWs and IAWs decrease with increasing electron-to-ion temperature ratio ($\mathrm{T_e/T_i}$) and normalized perpendicular electron inertial length ($ck_x/\omega_{pe}$). We evaluate the associated electric potentials and show that the electric potentials associated with KAWs are significantly affected in the high wavenumber ($k_x \rho_i$) regime, whereas the IAWs exhibit a decrease in electric potential along the magnetic field and an increase across it when the perpendicular electric field ($\mathrm{E_x}$) is enhanced. Additionally, we determine the net resonant speeds of particles in the perpendicular direction and demonstrate that these wave-particle interactions can efficiently heat the solar corona over extended distances (R\textsubscript{Sun}). Finally, we quantify the power transported by KAWs and IAWs through solar flux loop tubes, finding that both wave types deliver greater energy with increasing $\mathrm{T_e/T_i}$ and $ck_x/\omega_{pe}$. These insights not only deepen our theoretical understanding of wave-driven heating mechanisms but also provide valuable implications for interpreting solar wind, heliospheric, and magnetospheric dynamics.
\end{abstract}

\begin{keywords}
Sun -- Corona -- Heliosphere
\end{keywords}

%%%%%%%%%%%%%%%%%%%%%%%%%%%%%%%%%%%%%%%%%%%%%%%%%%

%%%%%%%%%%%%%%%%% BODY OF PAPER %%%%%%%%%%%%%%%%%%

\section{Introduction}

The understanding of solar wind acceleration began with Eugene Parker's groundbreaking work in 1956, where he proposed a mechanism that explains how the Sun's outer atmosphere, composed mainly of hydrogen plasma, could be accelerated to supersonic speeds \citep{parker1958dynamics}. This process results in a continuous outward flow of plasma from the Sun, known as the solar wind, which was later confirmed through direct observations by NASA's Mariner 2 spacecraft during its 1962 flyby of Venus \citep{hundhausen1968direct}. Despite significant progress, the precise mechanisms of energy dissipation in the solar wind and solar corona remain a topic of ongoing debate. Although heating the interplanetary collisionless plasma appears necessary to account for its nonadiabatic cooling \citep{richardson1995radial}, the specific kinetic processes involved are not yet fully understood. One possible mechanism for transferring energy from the lower solar atmosphere to the solar wind, thereby contributing to its acceleration, is the presence of hydromagnetic waves \citep{leer1982acceleration}. These waves can originate in either the solar corona or the lower solar atmosphere and propagate into the corona. Observational and theoretical studies support the idea that Alfv\'{e}n waves \citep{alfven1942existence} are the candidates for transporting energy through the solar atmosphere and raising temperatures in the coronal regions to millions of Kelvin. Support for the presence of Alfv\'{e}n waves has been provided by studies from \cite{de2007chromospheric}, \cite{okamoto2007coronal}, and \cite{cirtain2007evidence}.

In this paper, we explore the fundamental issue of coronal heating and solar wind acceleration by examining the role of low-frequency Alfv\'{e}n waves in both kinetic and inertial regimes. Recent decades have seen significant advances in our understanding of the solar atmosphere, fueled by an increase in observational data and the development of sophisticated analytical and numerical models of the solar wind \citep{coleman1968turbulence,belcher1971large,van2014alfven, kiyani2015dissipation,shiota2017turbulent}. \cite{osman2012kinetic} investigated the link between kinetic processes and intermittent turbulence, finding that inhomogeneous heating within the solar wind leads to significantly enhanced proton temperatures. Their findings show that this heating, evident in both the parallel and perpendicular components of the magnetic field, results in proton temperatures that are at least 3–4 times higher than those observed under typical, quiescent plasma conditions. Observations from spacecraft like Hinode have revealed that the energy contained in chromospheric magnetic field fluctuations, which propagate outward from the Sun, is sufficient to heat the solar corona and sustain its temperature at approximately 1 million Kelvin \citep{de2007chromospheric}. Furthermore, data from the Solar Dynamics Observatory (SDO) indicate that Alfv\'{e}n waves are pervasive in the transition region and low corona \citep{mcintosh2011alfvenic}. These findings highlight the critical need for developing a comprehensive three-dimensional model of the solar corona and inner heliosphere that incorporates Alfv\'{e}n wave turbulence. Such a model is essential for evaluating whether turbulence-driven mechanisms can successfully reproduce the emission structures observed in extreme ultraviolet (EUV) imagery. One of the most promising theoretical frameworks is the turbulence transport and dissipation model introduced by \cite{zank2011transport}, which was subsequently extended by \cite{shiota2017turbulent} to a fully three-dimensional solar wind configuration. This extended model provides a complete set of self-consistent, coupled equations that effectively describe the evolution and transport of turbulence in both sub-Alfv\'{e}nic and super-Alfv\'{e}nic inhomogeneous flows, thereby offering deeper insight into the complex dynamics of wave-particle interactions and energy transfer in the solar environment.

The study of Alfv\'{e}n wave turbulence in the solar wind has a rich history, tracing back to the groundbreaking work of \cite{coleman1968turbulence}, who, using Mariner 2 data, identified the significance of turbulence near 1 AU. The earliest models integrating Alfv\'{e}n wave turbulence into the solar wind were introduced by \cite{belcher1971large} and \cite{alazraki1971solar}. Subsequent advancements included the development of two-dimensional global corona models by
\cite{usmanov2000global} and \cite{hu2003coronal}. A notable progression came from \cite{suzuki2006solar}, who constructed a self-consistent one-dimensional model extending from the photosphere to the inner heliosphere, incorporating wave reflection and mode conversion from Alfv\'{e}n to slow waves. This model was later expanded to two dimensions by \cite{matsumoto2012connecting} to account for turbulent cascades.

Alfv\'{e}n waves that propagate outward from the Sun encounter stratification gradients, leading to partial reflection and the creation of counter-propagating waves, as demonstrated by studies such as \cite{heinemann1980non}, \cite{leroy1980propagation}, \cite{matthaeus1999coronal}, and the counterparts by \cite{zank2018theory, zank2021turbulence}. These counter-propagating waves are crucial for initiating the classical incompressible cascade, a key mechanism in coronal heating \citep{velli1989turbulent}. In regions with strong magnetic field gradients, particularly near active regions, reflection intensifies, potentially enhancing dissipation and leading to increased extreme ultraviolet emissions. Numerous turbulence transport and dissipation models [e.g., \citep{matthaeus1999coronal, zank2018theory, zank2021turbulence,yalim2024mixing} and references therein] have been proposed to investigate the heating and energization processes in the solar wind and corona. In this paper, we aim to determine the heating and acceleration of energetic particles in the solar corona using a kinetic plasma theory model.

Alfv\'{e}n waves become dispersive when their cross-field wavelength is short enough to be comparable to the primary kinetic length scale of the plasma \citep{singh2019anisotropic}. These dispersive Alfv\'{e}n waves are categorized into two distinct classes based on this comparison: kinetic Alfv\'{e}n waves (KAWs) and inertial Alfv\'{e}n waves (IAWs). Specifically, KAWs \citep{hasegawa1976kinetic} are defined as dispersive Alfv\'{e}n waves where the ion gyroradius ($\rho_i$) is greater than the electron inertial length ($\lambda_e$), meaning $\rho_i > \lambda_e$. Due to their dispersive nature, KAWs experience rapid damping, which can significantly contribute to inhomogeneous coronal heating, as supported by studies such as those by \cite{de1994coronal}, \cite{voitenko1995anomalous,voitenko1996flare}, \cite{elfimov1996noninductive}, \cite{asgari2012model}, \cite{testa2014evidence}, and \cite{morton2015investigating}. The anisotropic turbulence of KAWs, studied by \cite{singh2019anisotropic}, has been highlighted as a leading mechanism for particle energization and heating in the solar corona. KAWs have been extensively investigated in various settings, including the solar wind, solar corona, fusion reactors, and other astrophysical environments \citep{cramer2011physics, wu2020kinetic}. These investigations encompass experimental studies, theoretical analyses, simulations, and practical applications in plasma heating and particle acceleration. Comprehensive insights into KAWs can be found in review articles by \cite{gekelman1999review}, \cite{stasiewicz2000small}, \cite{wu2004recent}, \cite{keiling2009alfven}, and \cite{zhao2010nonlinear}.

Observational analyses have shown that KAWs dissipate energy and contribute to plasma heating as they propagate \citep{wygant2000polar,wygant2002evidence, lysak2003kinetic, gershman2017wave}. These waves are crucial in the transport, heating, and acceleration processes within space and astrophysical environments. For example, \cite{tsiklauri2005particle} found that phase mixing of KAWs in the solar coronal plasma can effectively accelerate electrons, a phenomenon also observed in magnetospheric plasma regions \citep{genot1999study, genot2004alfven, mottez2006comment}. Additionally, \cite{mottezelectron2011} conducted numerical studies on the interaction of an isolated solitary KAW packet with a plasma density cavity, forming small-scale coherent electric structures. \cite{goertz1984kinetic} studied KAWs on an auroral field line and found that these waves play a significant role in the charged particles' acceleration process.

In addition to the kinetic limit of Alfv\'{e}n waves, inertial Alfv\'{e}n waves (IAWs) \citep{goertz1979magnetosphere} are defined as; dispersive Alfv\'{e}n waves where the ion gyroradius ($\rho_i$) is smaller than the electron inertial length ($\lambda_e$), meaning $\rho_i < \lambda_e$. \cite{swift2007simulation} demonstrated that a wave-electric field becomes significant when the perpendicular wavelength of shear Alfv\'{e}n waves matches the electron inertial length, enabling these inertial scale waves to accelerate electrons parallel to the magnetic field. For those seeking a deeper understanding of the acceleration, heating mechanisms, properties, and intricate physical behavior of IAWs, several key studies offer invaluable insights. Noteworthy among these are the works of \cite{lysak1993generalized}, \cite{thompson1996electron}, \cite{chaston2000alfven}, \cite{watt2005self}, \cite{shukla2009study}, \cite{blanco2011electron},\citep{ayaz2019dispersion, ayaz2020alfven}. These seminal papers provide a fascinating exploration of IAW dynamics, offering a rich perspective on their role in space plasma physics and making them essential reading for anyone interested in this field. Furthermore, IAWs are thought to contribute to particle acceleration in various space plasma systems, including the Earth’s ionosphere \citep{thompson1996electron}, Jovian flux tubes \citep{damiano2023electron}, and Ganymede’s footprint tail aurora \citep{szalay2020alfvenic}. \cite{thompson1996electron} incorporated energy conservation into their model by allowing particles, specifically electrons, to influence the wave through Landau damping. In our research, we adopt a similar approach, where both kinetic Alfv\'{e}n waves and inertial Alfv\'{e}n waves transfer their energy to resonant solar plasma particles via Landau damping. This mechanism plays a crucial role in understanding the energy dynamics within the solar corona and other astrophysical environments.

In this study, we aimed to address key questions: (a) how KAWs and IAWs transport energy in the perpendicular direction (i.e., to the ambient magnetic field lines), (b) how the electric potentials associated with KAWs and IAWs influence the particles in the solar corona, (c), how KAWs and IAWs transport power across the solar flux loop tube, and (d) how charged particles get energized and accelerate and heat the solar corona. We utilized a kinetic plasma theory model to investigate these phenomena, varying the electron-to-ion temperature ratio at different heights relative to the solar radius. This approach allowed us to explore the dynamics of energy transport and particle behavior within the solar corona.

\section{Mathematical Formalism}

\begin{figure}
    \centering
    \includegraphics[width=0.5\linewidth]{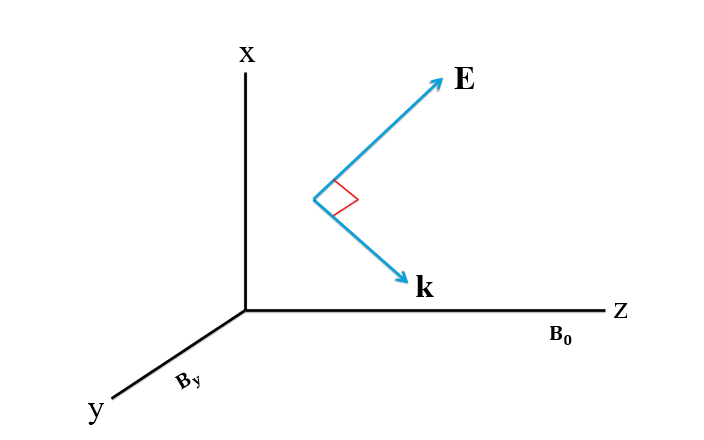}
    \caption{Schematic representation of the wave–field geometry in the x–z plane. The wave vector \textbf{k} and the perturbed electric field \textbf{E} are mutually perpendicular and lie within the x–z plane. The perturbed magnetic field $B_y$ of the wave, oriented in the y-direction, is perpendicular to both \textbf{k} and \textbf{E}, consistent with the right-hand rule for electromagnetic wave propagation. The uniform background magnetic field $B_0$ is aligned along the z-axis, defining the direction of the ambient magnetic field in the system.}
    \label{A}
\end{figure}

Employing the general Vlasov-Maxwell model, the dispersion relation for Alfv\'{e}n waves can be written as \citep{lysak1996kinetic,lysak1998relationship,khan2020solar}
 \begin{equation}
  \mathrm{D} = 
\left( 
\begin{array}{ccc}
\epsilon _{xx}-n_{\parallel}^2 &  & n_{\parallel} n_{\perp} \\ 
&  &  \\ 
n_{\parallel} n_{\perp} &  & \epsilon_{xx}-n_{\perp}^2%
\end{array}%
\right)
\label{eq:1},
\end{equation}
where $n_{\parallel,\perp}$ is the refractive index in the parallel and perpendicular directions, $\parallel$ and $\perp$ denote parallel and perpendicular, and $\epsilon_{xx,zz}$ is the permittivity tensor in the x and z components, respectively. The geometry of the system is illustrated in Figure \ref{A}, where the wave propagates obliquely in the x–z plane, characterized by the wavevector \textbf{k}. The ambient magnetic field $B_0$ is directed along the z-axis. The wave-induced perturbed magnetic field $B_y$ lies along the y-axis, while the associated electric field \textbf{E} is oriented along the x-axis. This configuration reflects the typical orthogonality between the electric and magnetic field perturbations in Alfv\'{e}n wave dynamics.

In Eqn. (\ref{eq:1}), $\epsilon_{xx}$ and $\epsilon_{xx}$ are given by \citep{lysak2003kinetic}
\begin{equation}
    \epsilon_{xx} = \frac{c^2}{V_{A}^2}\Bigr(1-\frac{3}{4} k_{x}^2 \rho_{i}^2 \Bigr)
    \label{eq:2},
\end{equation}
and 
\begin{equation}
    \epsilon_{zz} = -\Bigr(\frac{Z^{'} (\xi_e)}{2k_{z}^2 \lambda_{De}^2} +\frac{Z^{'} (\xi_i)}{2k_{z}^2 \lambda_{Di}^2}\Bigr)
    \label{eq:3}.
\end{equation}

Here $V_{A}$ is Alfv\'{e}n speed, $\rho_i$ is the gyroradius of ions, $c$ being the speed of light, $\lambda_{D e, i}$ is electron/ion Debye length, and $Z^{'}(\xi_{e, i})$ is the derivative of the plasma dispersion function 
\begin{equation*}
Z(\xi_{e,i})=\frac{1}{\sqrt{\pi}}\int_{-\infty}^{\infty}\frac{e^{-s^2}}{s-\xi_{e,i}}ds 
\end{equation*}
\citep{fried2015plasma} with argument $\xi_{e, i}=\omega/k_{z}v_{Te, i}$.

Eqn. (\ref{eq:3}) determines the limits of the waves, i.e., the kinetic or the inertial limits. In the next sections, we derive the governing expressions of both KAWs and IAWs.

\subsection{Kinetic Limits of Alfv\'{e}n waves (KAWs)}
In the kinetic limit (also called KAWs), $\xi_{e} \ll 1$ - hot electrons, and $\xi_{i} \gg 1$ - cold ions, the plasma dispersion function for electrons and ions can be written as
\begin{equation*}
Z^{'}(\xi_{e})\approx -2(1+i\xi_{e} \sqrt{\pi} e^{-\xi_{0e}^2}),
\end{equation*}
and
\begin{equation*}
Z^{'}(\xi_{i})\approx \xi_{0i}^{-2}-2i\xi_{0i}\sqrt{\pi} e^{-\xi_{0i}^2}.
\end{equation*}
On substituting $Z^{'}(\xi_{e,i})$ values in Eqn. (\ref{eq:3}) and back substituting Eqns. (\ref{eq:3}) and (\ref{eq:2}) in Eqn. (\ref{eq:1}) to get the real and imaginary frequencies of KAWs \citep{hasegawa1975kinetic, lysak1996kinetic, khan2020solar}; we obtain
\begin{equation}
    \omega_r=V_{A} k_{z}\Bigr[1+\frac{3}{4} k_{x}^2 \rho_{i}^2+\mathrm{\frac{T_e}{T_i}} k_{x}^2 \rho_{i}^2\Bigr]^{1/2}
    \label{eq:4},
\end{equation}
and
\begin{equation}
    \omega_i=-\frac{V_{A}^2 k_{z}}{2v_{te}}\mathrm{\frac{T_e}{T_i}} k_{x}^2 \rho_{i}^2\Bigr\{1+\Bigr(\mathrm{\frac{T_e}{T_i}}\Bigr)^{3/2} \sqrt{\frac{m_i}{m_e}} e^{-\xi_{0i}^2}\Bigr\}\sqrt{\pi}
    \label{eq:5}.
\end{equation}

\subsubsection{Poynting Flux Vector of KAWs:}

Following \cite{lysak2003kinetic}, the steady-state form of the Poynting theorem is: $\nabla \cdot \textbf{S}$ = -P, where \textbf{S} is the Poynting flux vector and P is the power dissipation rate given as
\begin{equation*}
    \textbf{S}=\text{Re} (E^{\ast} \times \textbf{B})/2 \mu_0,
\end{equation*}
and 
\begin{equation*}
    \mathrm{P}=\text{Re} (J^{\ast} \cdot \textbf{E})/2,
\end{equation*}
where $\mu_0$ is the permeability constant and \textbf{J} is the current density.

As shown in the geometry (Figure \ref{A}), the wave propagates in the x-z plane, hence, the y-component of the Poynting vector ($S_y$) is zero. We can write
\begin{equation}
    \nabla \cdot \textbf{S}= \frac{\partial}{\partial x} S_x (x,z)+\frac{\partial}{\partial z} S_z (x,z)
    \label{eq:6}
\end{equation}
\cite{ayaz2024solar}. In Eqn. (\ref{eq:6}), the first term on the right-hand side represents the perpendicular component of the Poynting flux, while the second term corresponds to its parallel component. It is well-established that the perpendicular wavevector ($k_x$) is significantly larger than the parallel wavevector ($k_z$), i.e., $k_x \gg k_z$, which directly implies that the energy flux along the x-direction ($S_x$) is much smaller than that along the z-direction ($S_z$). In the framework of ideal MHD and for low-frequency Alfv\'{e}n waves, energy is predominantly transported along the ambient magnetic field lines since the Poynting vector, given by $\delta \mathbf{E} \times \delta \mathbf{B}$, aligns with the background field $\mathrm{B_0}$. This behavior is captured by the first term in Eqn. (\ref{eq:6}).

However, in the kinetic (KAWs) and inertial (IAWs) regimes of Alfv\'{e}n waves, where dispersive and kinetic effects become significant, a non-negligible portion of energy can also propagate perpendicular to the magnetic field. This is primarily due to factors such as finite electron inertia, ion gyroradius effects, and parallel electric fields, and is described by the second term in Eqn. (\ref{eq:6}). While the parallel transport of energy has been extensively studied, the perpendicular contribution remains relatively unexplored, particularly in the solar coronal environment, where it could play a critical role in mechanisms like coronal loop heating.

Therefore, this study aims to thoroughly investigate the perpendicular energy transport and associated heating effects induced by Alfv\'{e}n waves in the solar corona. To facilitate this analysis, we decompose and evaluate the average contributions from both the x- and z-components in Eqn. (\ref{eq:6}) as follows:
\begin{equation}
    S_z= Re \Bigr(E^{\ast} B_y\Bigr)
    \label{eq:7},
\end{equation}
and 
\begin{equation}
    S_x= -\Bigr(\mathrm{\frac{E_z}{E_x}}\Bigr) S_z
    \label{eq:8}.
\end{equation}
Using Faraday's law, Eqn. (\ref{eq:7}) can be written as \citep{lysak2003kinetic} 
\begin{equation}
    \frac{\partial S_z}{\partial z}=k_{x} \text{Re} \Bigr(i \mathrm{\frac{E_z}{E_x}}\Bigr)S_z
    \label{eq:9}.
\end{equation}
In Eqns. (\ref{eq:8}) and (\ref{eq:9}), the ratio $E_z/E_x$ is obtained by solving the first row in 
\begin{equation*}
\left( 
\begin{array}{ccc}
\epsilon _{xx}-\frac{k_{z}^{2}c^{2}}{\omega ^{2}} &  & \frac{k_{x
}k_{z}c^{2}}{\omega ^{2}} \\ 
&  &  \\ 
\frac{k_{x }k_{z }c^{2}}{\omega ^{2}} &  & \epsilon _{zz}-\frac{%
k_{x }^{2}c^{2}}{\omega ^{2}}%
\end{array}%
\right) \left( 
\begin{array}{c}
\mathrm{E_{x}} \\ 
\\ 
\mathrm{E_{z}}%
\end{array}%
\right)=0
\label{eq:10},
\end{equation*}
\citep{lysak2003kinetic}. Thus, we have
\begin{equation}
    \mathrm{\frac{E_z}{E_x}}=\frac{c^2 k_{z}^2 -\omega^2 \epsilon_{xx}}{c^2 k_x k_z}
    \label{eq:10}
\end{equation}
Assuming $\omega=\omega_r +i \omega_i$ and that $\omega_i\ll \omega_r$, and considering the imaginary part, we have
\begin{equation}
    \text{Re} \Bigr(i \mathrm{\frac{E_z}{E_x}}\Bigr)=\frac{2 \epsilon_{xx} \omega_r \omega_i}{c^2 k_x k_z}
    \label{Eq:11}.
\end{equation}
Using Eqn. (\ref{Eq:11}) in Eqn. (\ref{eq:9}), yields
\begin{equation}
  \frac{\partial S_z}{\partial z}=\frac{2\epsilon_{xx} \omega_r \omega_i}{c^2 k_z}S_z
    \label{eq:12}, 
\end{equation}
whose solution is \citep{khan2020solar,ayaz2024asolar}
\begin{equation}
    S_z (z)=S(0) e^{\Bigr(\frac{2 \epsilon_{xx}\omega_r \omega_i}{c^2 k_z}z\Bigr)}.
    \label{eq:13}
\end{equation}

Back substitute Eqn. (\ref{eq:13}) in Eqn. (\ref{eq:8}) to get the solution for $S_x$ as
\begin{equation}
    S_x (z)=-\Bigr(\mathrm{\frac{E_z}{E_x}}\Bigr) S(0) e^{\Bigr(\frac{2 \epsilon_{xx}\omega_r \omega_i}{c^2 k_z}z\Bigr)}.
    \label{eq:14}
\end{equation}
The two Eqns. (\ref{eq:13}) and (\ref{eq:14}) describe the Poynting flux of KAWs, detailing the energy transport of these waves over distance. They quantify the electromagnetic energy transfer from the waves to the plasma as they propagate from the reference point $z = 0$. At $z = 0$, the waves are initially excited with a Poynting flux magnitude of $S(0)$. As the waves move forward, this Poynting flux gradually decreases.

\subsubsection{Electric Potential of KAWs:}

In recent research, \cite{zhang2022observations} followed the method established by \cite{lysak1996kinetic} to estimate the parallel electric potential for KAWs. The parallel electric potential ($\phi_z$) was expressed as:
\begin{equation}
    \mathrm{\phi_{z} \sim \lambda_z E_z}
    \label{eq:1a},
\end{equation}
where $\mathrm{\lambda_z}$ represents the wavelength or scale size in the direction parallel to the ambient magnetic field ($\mathrm{B_0}$), and $\mathrm{E_z}$ is the parallel component of the electric field. This formula provides an intuitive way to estimate the electric potential using the scale size of the wave and the parallel electric field.

We derived the parallel electric potential of KAWs using Eqn. (\ref{Eq:11}) and obtained a more detailed expression for $\mathrm{\phi_z}$. The parallel electric potential can be written as:
\begin{equation}
    \mathrm{\phi_{z}} \sim \frac{2 \pi}{k_z}\Bigr(\frac{2 \omega_r \omega_i \epsilon_{xx}}{c^2 k_x k_z}\Bigr) \mathrm{E_x}
    \label{eq:1b}.
\end{equation}
Eqn. (\ref{eq:1b}) shows that the parallel electric potential $\mathrm{\phi_z}$ is influenced by the complex wave frequency components, the dielectric properties of the medium, and the perpendicular electric field. The real ($\omega_r$) and imaginary ($\omega_i$) parts of the wave frequency play a crucial role in determining the wave's propagation and damping characteristics, directly affecting the parallel electric potential.

Similarly, in the perpendicular direction, we derive the perpendicular electric potential of KAWs as:
\begin{equation}
    \mathrm{\phi_{x}} \sim \frac{2 \pi}{k_x}\Bigr(\frac{c^2 k_x k_z}{2 \omega_r \omega_i \epsilon_{xx}}\Bigr) \mathrm{E_z}
    \label{eq:1c}.
\end{equation}
This expression (\ref{eq:1c}) highlights that the perpendicular electric potential $\mathrm{\phi_x}$ is related to the perpendicular wavenumber $k_x$ and the perpendicular dielectric tensor component ($\epsilon_{xx}$), along with the complex wave frequency. Unlike the parallel potential, $\mathrm{\phi_x}$ is also directly affected by the parallel electric field component $\mathrm{E_z}$, indicating the cross-field interaction between the wave's electric field and its associated electric potential in both directions.

These two expressions, (\ref{eq:1b}) and (\ref{eq:1c}), provide comprehensive formulae for understanding the electric potential of KAWs in both the parallel and perpendicular directions relative to the ambient magnetic field. The derived electric potentials are critical for understanding how KAWs transfer energy to charged particles in the plasma, leading to particle acceleration and heating. This is particularly important in space plasma environments like the solar corona, where such waves can efficiently contribute to the heating mechanisms and energy transport across magnetic field lines.

\subsubsection{Net Speed of the Particles:}
Following \cite{paraschiv2015physical}, the kinetic energy flux of the particles is given by
\begin{equation*}
    \mathrm{K_{energy}}  =\frac{\rho}{2}v^3,
\end{equation*}
where $v$ is the particles' resonant velocity and $\rho$ is the electrons mass density, respectively.

In our scenario involving KAWs, the wave energy per unit area per unit time is represented by the Poynting flux vector (\textbf{S}). From the law of conservation of energy, the kinetic energy expression  can also be written as
\begin{equation*}
    \mathrm{\textbf{S}}=\frac{\rho}{2}v^3,
\end{equation*}
which gives
\begin{equation}
    \mathrm{v}=\Bigr(\frac{2 S}{\rho}\Bigr)^{1/3}
    \label{eq:15}
\end{equation}
Eqn. (\ref{eq:15}) gives the speed the particles gained from the wave (i.e., KAWs). 

Particles that fulfill the resonant condition interact with the wave and gain energy when their velocity matches the wave's phase velocity. This velocity is the sum of the particles' initial velocity and the energy gained from the wave as it damps. Hence, we can write \citep{ayaz2024balfven, ayaz2025study}
\begin{equation}
    \mathrm{v_{r}}=\frac{\omega_r}{k_z}+\Bigr(\frac{2  \mathrm{\textbf{S}}}{\rho}\Bigr)^{1/3}
    \label{eq:16},
\end{equation}
where $\mathrm{v_{r}}$ is the net resultant velocity of resonant particles, the wave vector's parallel component ($k_z$) is considered in the equation because these waves mostly deliver their energy along the magnetic field lines.

In the kinetic limits of Alfv\'{e}n waves, we are interested in the contribution of charged particle heating and acceleration in the perpendicular direction that is across the magnetic field lines. However, we can proceed with the perpendicular Poynting flux expression (\ref{eq:14}). Substituting Eqn. (\ref{eq:14}) in Eqn. (\ref{eq:16}) yields the perpendicular net velocity of the perpendicular propagating particles as
\begin{equation}
 \mathrm{v_{rp}}=\frac{\omega_r}{k_z}-\Bigr[\frac{2}{\rho}\Bigr(\mathrm{\frac{E_z}{E_x}}\Bigr) S(0) e^{(\frac{2 \epsilon_{xx}\omega_r \omega_i}{c^2 k_z}z)}\Bigr]^{1/3}
    \label{eq:17}
\end{equation}
Eqn. (\ref{eq:17}) gives the resultant resonant speed of the particles in a perpendicular direction.

\subsubsection{The Power transfer rate of KAWs}

Our focus in this study is also the heating and energization of the solar corona (particularly flux tube loops). How KAWs transport energy and propagate in the flux tube is still under active debate in the solar corona. The previously derived expressions (\ref{eq:13} and \ref{eq:14}) of the Poynting flux vectors in both the parallel and perpendicular directions give an estimate of this element. Using the approach we adopted in our previous papers \citep{ayaz2024asolar,ayaz2024balfven, ayaz2025study}, we calculate the total power transfer rate of KAWs by integrating Eqns. (\ref{eq:13}) and (\ref{eq:14}) from 0 to $\pi$ and $\pi/2$. Hence, Eqn. (\ref{eq:13}) gives
\begin{equation}
    \mathrm{P_z}=\frac{S_z (z)}{S (0)}=a^2 \Bigr(e^{(\frac{\epsilon_{xx} \omega_r \omega_i h \pi}{c^2 k_z})}\Bigr)\pi
    \label{eq:18}.
\end{equation}
Similarly, expression (\ref{eq:14}) yields
\begin{equation}
    \mathrm{P_x}=\frac{S_x (z)}{S (0)}=-\frac{a c^2 k_z}{\epsilon_{xx}\omega_r \omega_i}\Bigr(\frac{E_z}{E_x}\Bigr) \Bigr[1-e^{(\frac{\epsilon_{xx}\omega_r \omega_i h \pi}{c^2 k_z})} \Bigr]\pi
    \label{eq:19}.
\end{equation}
Eqns. (\ref{eq:18}) and (\ref{eq:19}) give the power transfer by KAWs in the corresponding parallel and perpendicular directions in the flux tube loop. The ratio of these expressions yields the total power transfer rate of KAWs in the tube loop, given by
\begin{equation}
    \mathrm{P_t}=\frac{\mathrm{P_x}}{\mathrm{P_z}}=-\frac{c^2 k_z}{a\epsilon_{xx}\omega_i \omega_r \times e^{(\frac{\epsilon_{xx}\omega_r \omega_i h \pi}{c^2 k_z})}} \Bigr(\mathrm{\frac{E_z}{E_x}}\Bigr)\Bigr[1-e^{(\frac{\epsilon_{xx}\omega_r \omega_i h \pi}{c^2 k_z})}\Bigr]
    \label{eq:20}.
\end{equation}
In the above expression (\ref{eq:20}), $h$ is the height and $a$ is the radius of the cross-section of the flux tube, which we assumed to be circular, the details of which are given in our recent paper \citep{ayaz2025study}. Following \cite{li2023modeling}, we take $a$ in the range of 700 – 7000 km, in the numerical evaluation. This expression gives the total power transfer rate of KAWs in the flux tube loop.

\subsection{Inertial Limits of Alfv\'{e}n waves (IAWs)}
For inertial limits (i.e., IAWs), both electrons and ions are cold, i.e., $\xi_{e, i} \gg 1$. Thus, the plasma dispersion function for ions and electrons, as given in section 2.1, is
\begin{equation*}
    Z^{'}(\xi_{e,i})\approx (\xi_{e,i}^{2}-2i\xi_{e,i} \sqrt{\pi} e^{-\xi_{e,i}^2}).
\end{equation*}
Using values of the $Z$-function in Eqn. (\ref{eq:3}) and substituting Eqns. (\ref{eq:3}) and (\ref{eq:2}) back in Eqn. (\ref{eq:1}), we obtain the real and imaginary frequencies for IAWs as \citep{shukla2009study}
\begin{equation}
    \omega_{ri}^2=\frac{k_{z}^2 V_{A}^2}{1+\frac{k_{x}^2 c^2}{\omega_{pe}^2}}
    \label{eq:21},
\end{equation}
and 
\begin{equation}
    \omega_{ii}=-\frac{k_{x}^2 k_{z} V_{A}^4 c^2 exp(-\xi_{0e}^2)}{\omega_{pe}^2 v_{te}^3 \Bigr( 1+\frac{k_{x}^2 c^2}{\omega_{pe}^2}\Bigr)^{3}} \sqrt{\pi}
    \label{eq:22}.
\end{equation}
It should be noted that the additional subscript \textquoteleft i\textquoteright in the above expressions is used to differentiate the dispersion relations of IAWs from the kinetic limit of Alfv\'{e}n waves.

\subsubsection{Poynting Flux Vector of IAWs:}

Having established the dispersion relations given by Eqns. (\ref{eq:21}) and (\ref{eq:22}), calculations of the Poynting flux vectors for IAWs become a matter of straightforward algebraic manipulation. By applying similar algebraic techniques as before, we can determine the solutions for both the parallel and perpendicular components of the Poynting flux vectors, denoted as $S_x$ and $S_z$, for IAWs:
\begin{equation}
   S_{xi} (z)=-\Bigr(\mathrm{\frac{E_z}{E_x}}\Bigr) S(0) e^{\Bigr(\frac{2 \epsilon_{xx}\omega_{ri} \omega_{ii}}{c^2 k_z}z\Bigr)}
   \label{eq:23},
\end{equation}
and 
\begin{equation}
    S_{zi} (z)=S(0) e^{\Bigr(\frac{2 \epsilon_{xx}\omega_{ri} \omega_{ii}}{c^2 k_z}z\Bigr)}
    \label{eq:24}, 
\end{equation}
with 
\begin{equation*}
    \mathrm{\frac{E_z}{E_x}}=\frac{2 \epsilon_{xx} \omega_{ri} \omega_{ii}}{c^2 k_x k_z},
\end{equation*}
respectively. Eqns. (\ref{eq:23}) and (\ref{eq:24}) represent the Poynting flux vector for IAWs in both the perpendicular and parallel directions. The analytical calculations for these expressions was performed similarly to the approach used for KAWs.

\subsubsection{Electric Potential of IAWs:}

Using a similar approach as applied to KAWs, the parallel electric potential $\mathrm{\phi_{zi}}$ of IAWs is derived as:
\begin{equation}
    \mathrm{\phi_{zi}} \sim \frac{2 \pi}{k_z}\Bigr(\frac{2 \omega_{ri} \omega_{ii} \epsilon_{xx}}{c^2 k_x k_z}\Bigr) \mathrm{E_x}
    \label{eq:1d}.
\end{equation}
Equation (\ref{eq:1d}) expresses the parallel electric potential of IAWs, showing that it depends on the wave’s frequency components, dielectric properties of the plasma, and electric field in the perpendicular direction. The presence of both $\omega_{ri}$ and $\omega_{ii}$ underscores the role of wave damping and growth processes in determining the strength of the parallel electric potential, similar to KAWs.

In the perpendicular direction, the perpendicular electric potential $\mathrm{\phi_{xi}}$ of IAWs is given by:
\begin{equation}
    \mathrm{\phi_{xi}} \sim \frac{2 \pi}{k_x}\Bigr(\frac{c^2 k_x k_z}{2 \omega_{ri} \omega_{ii} \epsilon_{xx}}\Bigr) \mathrm{E_z}
    \label{eq:1e}.
\end{equation}
This equation reveals how the perpendicular electric potential is influenced by the perpendicular wavenumbers, the plasma's dielectric properties, and the complex frequency of the waves. The dependence on $\mathrm{E_z}$ highlights the cross-field interaction between the wave's electric potential and the electric fields in both parallel and perpendicular directions.

The two expressions (\ref{eq:1d}) and (\ref{eq:1e}) offer a comprehensive framework for understanding the electric potential of IAWs concerning the ambient magnetic field ($\mathrm{B_0}$). These potentials play a critical role in the wave-particle interactions, contributing to particle acceleration and heating in the plasma. The parallel and perpendicular electric potentials govern how energy is transferred from IAWs to charged particles, influencing their dynamics in space plasma environments such as the solar wind and magnetospheric regions.

The distinct behavior of the parallel and perpendicular electric potentials highlights the importance of wave geometry and plasma conditions in determining the effectiveness of energy transfer. This knowledge is crucial for understanding the mechanisms behind particle acceleration and heating in regions where inertial effects dominate, as seen in IAWs. By exploring these potentials, we can gain deeper insight into the energy dynamics within plasma systems affected by inertial Alfv\'{e}n waves.

\subsubsection{Net Speed of the Particles:}

Expressions (\ref{eq:23}) and (\ref{eq:24}) provide a basis for calculating the net speed of particles resulting from wave-particle interactions (this time, IAWs interact with particles). Specifically, these equations allow us to determine both the parallel and perpendicular components of the resultant particle velocities after they interact with the waves (IAWs). Evaluate Eqn. (\ref{eq:16}) for IAWs (i.e., substitute Eqns. \ref{eq:21} and \ref{eq:24}) to get the net parallel speed of particles as
\begin{equation}
     \mathrm{V_{ni}}=\frac{\omega_{ri}}{k_z}+\Bigr[\frac{2}{\rho} S(0) e^{(\frac{2 \epsilon_{xx}\omega_{ri} \omega_{ii}}{c^2 k_z}z)}\Bigr]^{1/3}
    \label{eq:25}.
\end{equation}
 Similarly, substitute Eqns. (\ref{eq:21}) and (\ref{eq:23}) in Eqn. (\ref{eq:16}) yields the resultant perpendicular speed of the particles given by
\begin{equation}
     \mathrm{V_{nip}}=\frac{\omega_{ri}}{k_z}-\Bigr[\frac{2}{\rho} \Bigr(\mathrm{\mathrm{\frac{Ez}{Ex}}}\Bigr) S(0) e^{(\frac{2 \epsilon_{xx}\omega_{ri} \omega_{ii}}{c^2 k_z}z)}\Bigr]^{1/3}
    \label{eq:26}.
\end{equation}
In the above expressions, Eqn. (\ref{eq:25}) is evaluated in our previous paper, for instance, see \cite{batool2024acceleration}, however, the counterpart, i.e., Eqn. (\ref{eq:26}) has not yet been explored in the existing literature. Our focus is on evaluating the perpendicular resultant speed of particles in this work.

\subsubsection{The Total Power Transfer Rate of IAWs}

By doing the same calculations as we did previously to calculate the power rate of KAWs, the total power rate ($\mathrm{P_{ti}}$) of IAWs can be found by taking the integration of Eqns. (\ref{eq:23}) and (\ref{eq:24}), hence the ratio is
\begin{equation}
    \mathrm{P_{ti}}=-\frac{c^2 k_z}{\omega_{ii} \omega_{ri}\epsilon_{xx} a e^{(\frac{\epsilon_{xx}\omega_{ri} \omega_{ii} h \pi}{c^2 k_z})}} \Bigr(\mathrm{\mathrm{\frac{E_z}{E_x}}}\Bigr)\Bigr[1-e^{(\frac{\epsilon_{xx}\omega_{ri} \omega_{ii} h \pi}{c^2 k_z})}\Bigr]
    \label{eq:27}.
\end{equation}
Eqn. (\ref{eq:27}) represents the total power transfer rate of IAWs in the solar flux loop tube. The contribution of the electric field ($\mathrm{E_z/E_x}$) should be noted because the wave delivers power across the loop, where the electric field pointing along the x-axis is also influencing the power.

\section{Results and Discussion}

This study examined KAWs and IAWs within the solar coronal region. The parameters used for our analysis include a temperature range of $\sim \times 10^6$ Kelvin \citep{de2015recent}, the magnetic field strength is (50 - 100) Gauss \citep{zirin1996mystery, gary2001plasma}, the wavenumber ratio $k_x/k_z$ spanning from 100 to 115 \citep{chen2012kinetic}, and the particle density on the order of $\sim 5\times 10^{9}$ cm\textsuperscript{-3} \citep{pneuman1973solar}. These parameter values were chosen to accurately represent the conditions in the solar corona and ensure a comprehensive analysis of wave behavior in this environment.

\subsection{Analysis of KAWs}
\begin{figure}
        \centering
       {
            \includegraphics[width=.47\linewidth]{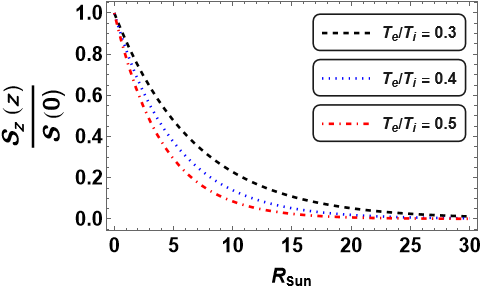}
                    }\quad
        {
            \includegraphics[width=.47\linewidth]{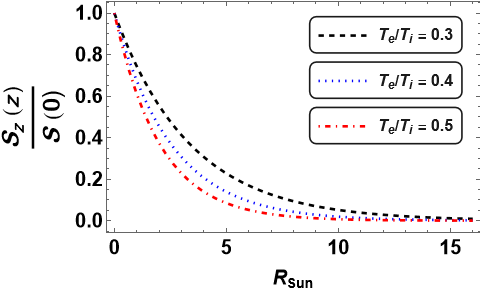}
                    }
        \caption{Normalized Poynting flux vector ($S_z (z)/S(0)$) of KAWs plotted against distance in solar radii R\textsubscript{Sun}, illustrating energy propagation in the parallel direction. The plot is generated using Eqn. (\ref{eq:13}) with the following parameters: Alfv\'{e}n speed $V_{A} \approx 1.85\times 10^{8}$ cm/s, electron thermal speed $v_{te} \approx 1.34\times 10^{9}$ cm/s, ion thermal speed $v_{ti} \approx 1.9\times 10^{7}$ cm/s, $\rho_i \approx 40$ cm, perpendicular wavenumber $k_x \rho_i \approx 0.02$, perpendicular wavenumber $k_x \approx 5.8\times 10^{-3}$ cm\textsuperscript{-1}, and parallel wavenumber $k_z \approx 5.8\times 10^{-10}$ cm\textsuperscript{-1}, respectively. The left panel shows the flux rate for height h = 0.05 R\textsubscript{Sun}, while the right panel corresponds to h = 0.1 R\textsubscript{Sun}. The flux rate is evaluated across different electron-to-ion temperature ratios $\mathrm{T_e/T_i}$, showing the impact of these variations on energy dissipation.}
    \label{B}
\end{figure}

In our investigation of KAWs, we focus primarily on their perpendicular energy transport and heating effects. However, for comparative analysis, we also examine parallel heating (e.g., the parallel Poynting flux vector $S_z$), as illustrated in Figure \ref{B}. Our findings reveal that the Poynting flux vector of KAWs is notably influenced by variations in the electron-to-ion temperature ratio ($\mathrm{T_e/T_i}$).

At a height of h = 0.05 R\textsubscript{Sun} (left panel), KAWs effectively heat the coronal region over a longer distance. In contrast, at h = 0.1 R\textsubscript{Sun} (right panel), the normalized parallel Poynting flux ($S_z (z)/S(0)$) of KAWs dissipates more rapidly over shorter distances R\textsubscript{Sun}. In the parallel directions, KAWs carry most of the energy and can heat the corona for larger distances. As $\mathrm{T_e/T_i}$ increases, $S_z (z)/S(0)$ dissipates faster, implying that the strong parallel electric fields associated with the wave lead to more efficient energy transport/transfer from wave to particles. This results in faster dissipation of wave energy to the particles through the Landau damping mechanism. These observations align with previous studies by \cite{lysak2003kinetic,khan2019distinct,khan2020solar} and \cite{lysak2023kinetic} who also explored the Poynting flux characteristics of KAWs and found similar trends for the $S_z (z)/S(0)$.

In the kinetic regime of Alfv\'{e}n waves, where $k_x \gg k_x$, KAWs predominantly carry energy in the parallel direction, with only a minimal amount of energy being transported perpendicularly. Recent research by \cite{lysak2023kinetic, ayaz2024asolar} suggests that perpendicular heating becomes significant when KAWs experience rapid dissipation.

\begin{figure}
        \centering
       {
            \includegraphics[width=.47\linewidth]{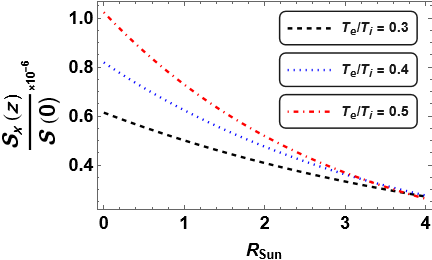}
                    }\quad
        {
            \includegraphics[width=.47\linewidth]{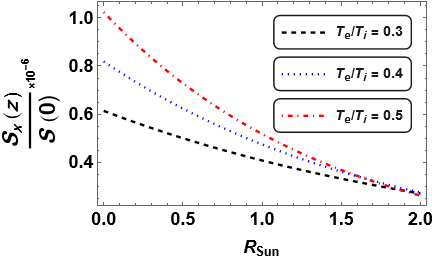}
                    }
        \caption{Normalized perpendicular Poynting flux vector $S_x (z)/S(0)$ of KAWs as a function of distance in solar radii R\textsubscript{Sun}, illustrating the energy transport across different regions of the solar corona. These plots are derived from Eqn. (\ref{eq:14}), using the same parameters as in Figure \ref{B}. The left panel corresponds to a height of h = 0.05 R\textsubscript{Sun}, and the right panel represents h = 0.1 R\textsubscript{Sun}. The Poynting flux is calculated for various electron-to-ion temperature ratios $\mathrm{T_e/T_i}$, showing how temperature differences affect energy dissipation over distance.}
    \label{C}
\end{figure}

Figure \ref{C}. depicts the normalized perpendicular Poynting flux vector ($S_x (z)/S(0)$) of KAWs for various electron-to-ion temperature ratios ($\mathrm{T_e/T_i}$). The results show that $S_x (z)/S(0)$ increases with higher $\mathrm{T_e/T_i}$ values. Specifically, at a height of h = 0.05 R\textsubscript{Sun} (left panel), the KAWs are observed to heat the corona over a larger distance compared to h = 0.1 R\textsubscript{Sun} (right panel) for the same $\mathrm{T_e/T_i}$ values. This indicates that KAWs dissipate more rapidly in the perpendicular direction, enhancing the heating effect in regions closer to the Sun. A higher $\mathrm{T_e/T_i}$ enhances the electronic response, which in turn amplifies $\mathrm{E_x}$, allowing part of the parallel Poynting flux (i.e., the energy transported along the magnetic field) to contribute to perpendicular plasma heating. This is a subtle - yet important mechanism in coronal heating and wave-particle energy transfer, especially in the KAWs and IAWs regimes.

\begin{figure}
    \centering
    \includegraphics[width=0.5\linewidth]{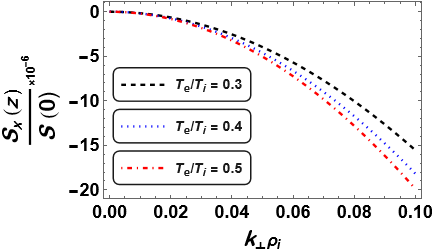}
    \caption{Normalized perpendicular Poynting flux vector $S_x (z)/S(0)$ of KAWs plotted against the normalized wavenumber $k_x \rho_i$. The flux $S_x (z)/S(0)$ is computed for various $\mathrm{T_e/T_i}$ values, with the height fixed at h = 0.05 R\textsubscript{Sun}. This analysis demonstrates how changes in the temperature ratio influence the energy dissipation behavior as a function of wavenumber in the solar corona.}
    \label{D}
\end{figure}

We examined the normalized $S_x (z)/S(0)$ of KAWs as a function of the normalized wavenumber $(k_x \rho_i)$ for varying $\mathrm{T_e/T_i}$, as illustrated in Figure \ref{D}. The results indicate that $S_x (z)/S(0)$ decays more rapidly at higher $\mathrm{T_e/T_i}$ values. When the electron-to-ion temperature is lower (e.g., $\mathrm{T_e/T_i} \approx 0.3$), KAWs effectively heat the coronal region over a broader wavenumber range before fully dissipating.

Furthermore, we assessed $S_x (z)/S(0)$ at different heights relative to the Sun's radius but observed no significant height impact on the dissipation rate. Whether at h = 0.1 or h = 0.05 R\textsubscript{Sun}, the rate at which $S_x (z)/S(0)$ decays remains consistent, as reflected in the overlap of the curves. This suggests that height does not notably influence the flux dissipation in the specific wavenumber range under consideration. Consequently, we excluded the plot for h = 0.1 R\textsubscript{Sun}, focusing instead on the uniform dissipation behavior observed across different heights.

\begin{figure}
        \centering
       {
            \includegraphics[width=.47\linewidth]{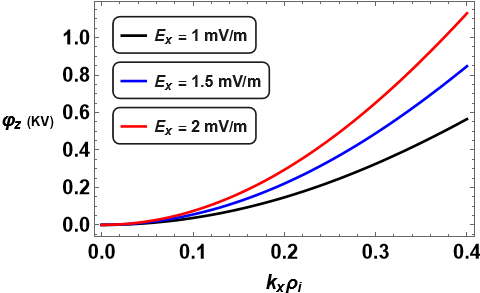}
                    }\quad
        {
            \includegraphics[width=.47\linewidth]{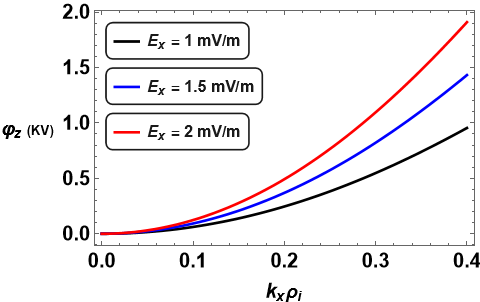}
            }
        \caption{The electric potential $\phi_z$ of KAWs in the parallel direction verses normalized wavenumber $k_x \rho_i$. The curves are generated using Eq. (\ref{eq:1b} for different values of $E_x$, and the other parameters are the same as in Figure (\ref{B}). In the left panel, $\mathrm{T_e/T_i}$ = 0.3, and $\mathrm{T_e/T_i}$ = 0.5 in the right panel, respectively.}
    \label{E1}
\end{figure}

Figure \ref{E1} presents the parallel electric potential $\mathrm{\phi_z}$ associated with KAWs. The plot demonstrates a clear trend: as the perpendicular wavenumber $k_x \rho_i$ increases, $\mathrm{\phi_z}$ is significantly enhanced, particularly when the perpendicular electric field $\mathrm{E_x}$ increases. This indicates that larger perpendicular wave scales, in combination with strong electric fields, lead to a notable rise in the parallel electric potential. Such an increase in $\mathrm{\phi_z}$ suggests stronger wave-particle interactions, which can result in more efficient energy transfer to the charged particles in the plasma, contributing to both particle acceleration and heating. 

Additionally, we evaluated $\mathrm{\phi_z}$ for different $\mathrm{T_e/T_i}$ values to explore how temperature variation influences the potential. In the left panel of Figure \ref{E1}, where $\mathrm{T_e/T_i} = 0.3$, the magnitude of $\mathrm{\phi_z}$ is approximately 1 kV. As we increase the temperature ratio to $\mathrm{T_e/T_i} = 0.5$ (right panel), $\mathrm{\phi_z}$ rises to around 2 kV. This change highlights the sensitivity of the parallel electric potential to variations in plasma temperature, especially the relative temperature of electrons and ions. A higher electron temperature relative to ions increases the parallel electric potential, potentially leading to more effective energy transfer from the waves to the particles. This imbalance strengthens the wave-particle interaction, accelerating electron beams to higher energies. As a result, electrons gain substantial energy, contributing to more intense heating and acceleration within the plasma. This process plays a key role in generating high-energy electrons, which are crucial for understanding particle dynamics in regions like the solar corona and solar wind.

Recently, \cite{zhang2022observations} analyzed the parallel potential of KAWs and reported that $\mathrm{\phi_z}$ can reach several KV in their model, with specific findings of $\mathrm{\phi_z} \sim 1.66 \mathrm{kV}$. Similarly, \cite{kostarev2021alfven} studied the parallel electric field associated with Alfv\'{e}n waves and found that the parallel electric field potential can reach values in the kV range, which is sufficient to accelerate precipitating auroral electrons. Their findings underline the significance of strong electric potentials in wave-driven particle acceleration, particularly in auroral regions where such waves are prominent.

In our results, the parallel electric potential of KAWs in the chosen solar coronal region also reaches several kilovolts, aligning well with these observations. This agreement reinforces the validity of our model and its relevance to understanding KAWs' behavior in the solar corona. The high $\mathrm{\phi_z}$ values observed suggest that KAWs could play a crucial role in driving particle acceleration and heating in the corona, particularly under conditions of varying electron-to-ion temperature ratios. This potential contributes to the broader energy dynamics of the solar atmosphere, where KAWs help transfer energy from the wave fields into the particle populations, influencing the overall heating of the solar corona.

\begin{figure}
        \centering
       {
            \includegraphics[width=.47\linewidth]{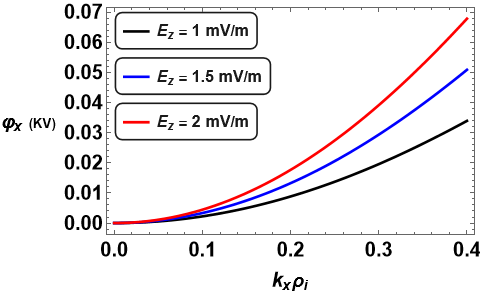}
                    }\quad
        {
            \includegraphics[width=.47\linewidth]{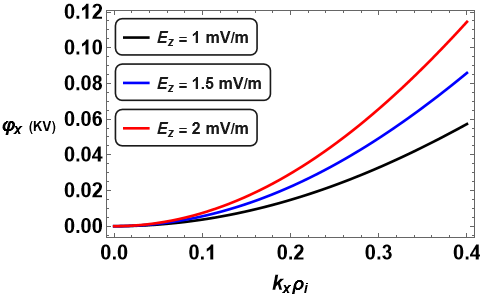}
            }
        \caption{The electric potential $\mathrm{\phi_x}$ of KAWs in the perpendicular direction as a function of the normalized wavenumber $k_x \rho_i$ for different values of $E_z$. The curves are generated using Eq. (\ref{eq:1c} with the same parameter values as in Figure (\ref{B}). In the left panel, $\mathrm{T_e/T_i}$ = 0.3, and $\mathrm{T_e/T_i}$ = 0.5 in the right panel, respectively.}
    \label{E2}
\end{figure}

We also investigate the behavior of the perpendicular electric potential $\mathrm{\phi_x}$ associated with KAWs, particularly in regimes where dispersive and kinetic effects become significant. Our analysis reveals that $\mathrm{\phi_x}$ exhibits a marked increase at relatively higher values of the normalized wavenumber $k_x \rho_i$, see Figure \ref{E2}. This enhancement becomes especially prominent in regions where the parallel electric field component $\mathrm{E_z}$ is also strong, highlighting a direct coupling between parallel field dynamics and perpendicular potential structures. Physically, this implies that as the wave becomes increasingly oblique (i.e., larger $k_x \rho_i$), the wave–particle interaction strengthens in the perpendicular direction. The rapid dissipation of wave energy across the magnetic field lines leads to a redistribution of energy, favoring enhanced perpendicular heating and energy transport. Consequently, the perpendicular electric potential $\mathrm{\phi_x}$ becomes more dominant, playing a crucial role in energizing particles perpendicular to the ambient magnetic field in high-$k_x \rho_i$ regimes. These findings emphasize the importance of perpendicular kinetic scales in modulating wave energy dissipation and suggest that $\mathrm{\phi_x}$ could serve as a key diagnostic for identifying localized heating sites and turbulent dissipation regions in the solar corona and solar wind.

As depicted in Figure \ref{E2}, it is evident that the magnitude of $\phi_x$ is substantially lower than that of the parallel electric potential. This difference is physically intuitive, given that KAWs primarily transport energy and heat particles along the magnetic field lines, meaning in the parallel direction. As noted by \cite{lysak2003kinetic} and \cite{khan2020solar}, KAWs are much more effective in driving parallel heating, particularly in environments like the aurora, where charged particles are confined by strong magnetic fields.

In contrast, only a small portion of the wave energy is transferred in the perpendicular direction, across the magnetic field lines. The reason for this is the rapid dissipation rate of KAWs in the perpendicular direction. As \cite{ayaz2024asolar} discussed, the wave energy in this direction quickly dissipates due to strong wave damping, which limits the distance over which energy can be transported perpendicularly. The fast dissipation of KAWs across magnetic field lines also means that less energy is available to accelerate particles in the perpendicular direction, contributing to the lower $\phi_x$ values.

A more detailed analysis of this dissipation was presented in our earlier discussion of the Poynting flux, where it was shown that energy transported in the parallel direction persists over longer distances, while energy in the perpendicular direction dissipates rapidly. Figure \ref{E2} visually supports this conclusion, illustrating that the perpendicular potential of KAWs is smaller in magnitude compared to the parallel potential.

It is important to emphasize that variations in the electron-to-ion temperature ratio (as $\mathrm{T_e/T_i}$) significantly affect the perpendicular electric potential, much like their influence on the parallel electric potential. Specifically, as as $\mathrm{T_e/T_i}$ increases, both the parallel ($\mathrm{\phi_z}$) and perpendicular ($\mathrm{\phi_x}$) electric potentials exhibit a corresponding rise. This trend indicates that although the perpendicular potential is generally smaller in magnitude, it is nonetheless highly responsive to changes in plasma temperature conditions. Such sensitivity implies that enhanced electron thermal energy can indirectly amplify perpendicular energy transfer mechanisms, thereby affecting particle acceleration and localized heating across magnetic field lines. The observed rise in $\mathrm{\phi_x}$ may stem from nonlinear coupling effects, where strong parallel heating — driven by elevated $\mathrm{T_e}$ — modifies the overall wave dynamics.

\begin{figure}
        \centering
       {
            \includegraphics[width=.47\linewidth]{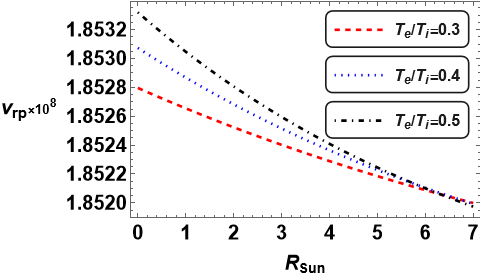}
                    }\quad
        {
            \includegraphics[width=.47\linewidth]{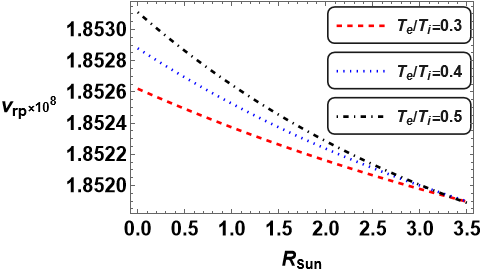}
            }
        \caption{Net resonant speed $\mathrm{V_{rp}}$ of particles in the perpendicular direction as a function of distance in solar radii $R_{Sun}$. 
        The plots are derived from Eqn. (\ref{eq:17}) using the same parameters as in Figure (\ref{B}). The left panel represents h = 0.05 R\textsubscript{Sun}, and the right panel corresponds to h = 0.1 R\textsubscript{Sun}. The resonant speed $V_{rp}$ is evaluated for different electron-to-ion temperature ratios $\mathrm{T_e/T_i}$, showing a significant enhancement in $\mathrm{V_{rp}}$ magnitude with increasing $\mathrm{T_e/T_i}$ values.}
    \label{E}
\end{figure}

Building on the investigation of KAWs and their impact on perpendicular heating, we examined the net perpendicular speed of particles as they gain energy from these waves, leading to plasma heating and acceleration. Figure \ref{E} illustrates the resultant perpendicular velocity $\mathrm{V_{rp}}$ of particles for various values of $\mathrm{T_e/T_i}$. We observed that the magnitude of $\mathrm{V_{rp}}$ increases with higher $\mathrm{T_e/T_i}$ values, indicating enhanced particle acceleration and heating.

In the region close to the Sun’s surface, specifically at a height of h = 0.05 R\textsubscript{Sun} (left panel), particles can heat the plasma over a significantly larger distance compared to the region at h = 0.1 R\textsubscript{Sun} (right panel). This suggests that in smaller-height regions, such as 0.05 R\textsubscript{Sun}, particles are more effectively accelerated and can transfer energy to the surrounding plasma over greater distances, approximately up to 7 solar radii (R\textsubscript{Sun}). This enhanced heating and acceleration underscore the critical role of KAWs in the dynamics of solar corona and solar wind regions, where they facilitate efficient energy transfer and plasma heating.

\begin{figure}
    \centering
    \includegraphics[width=0.5\linewidth]{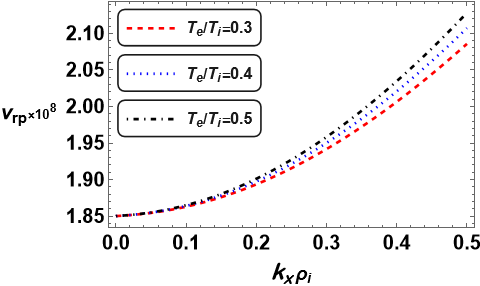}
    \caption{Net resonant speed $\mathrm{V_{rp}}$ of particles as a function of the normalized wavenumber $k_x \rho_i$. $S_z (x)/S(0)$. The evaluation is performed for various electron-to-ion temperature ratios $\mathrm{T_e/T_i}$ at a fixed height of h = 0.05 R\textsubscript{Sun}. This analysis provides insights into how the $\mathrm{V_{rp}}$ behavior of KAWs varies with different temperature ratios, particularly with the normalized wavenumber.}
    \label{F}
\end{figure}

We also explored the resultant perpendicular resonance speed of KAWs across a range of different electron-to-ion temperatures and the normalized wavenumber $k_x \rho_i$ values. As shown in Figure \ref{F}, $\mathrm{V_{rp}}$ is significant at larger $k_x \rho_i$ values. In contrast, at smaller $k_x \rho_i$, the curves converge, showing minimal dependence on temperature, which suggests that the influence of $\mathrm{T_e/T_i}$ diminishes in these conditions.

Interestingly, the net speed of the particles $\mathrm{V_{rp}}$ escalates with a higher $\mathrm{T_e/T_i}$ and larger $k_x \rho_i$ values, leading to enhanced particle acceleration and heating in the perpendicular direction within the solar corona. This behavior indicates that while temperature plays a crucial role in the dynamics of KAWs, its effect is primarily noticeable when coupled with significant wavenumber values. Thus, KAWs facilitate effective energy transfer, contributing to the coronal heating process, especially under conditions of higher temperatures and larger wavenumber scales.

\begin{figure}
        \centering
       {
            \includegraphics[width=.47\linewidth]{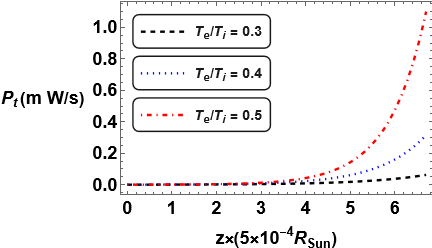}
                    }\quad
        {
            \includegraphics[width=.47\linewidth]{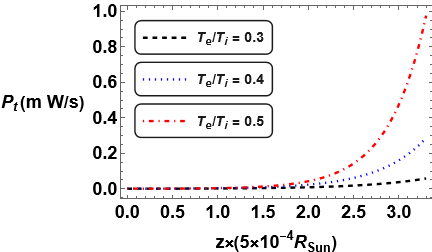}
            }
        \caption{The total power transfer rate ($\mathrm{P_t}$) of KAWs versus normalized distance $z\times (0.0005$ R\textsubscript{Sun}) for different values of $\mathrm{T_e/T_i}$. We assumed the cross-section $a \approx \times 10^{7}$ cm, and the other parameters are the same as those we used in Figure \ref{B}. h = 0.05 R\textsubscript{Sun} in the left panel and 0.1 R\textsubscript{Sun} in the right panel, respectively.}        
    \label{GP}
\end{figure}

Figure \ref{GP} illustrates the total power delivered by KAWs within the flux tube. The plots reveal a clear enhancement in power transfer as $\mathrm{T_e/T_i}$ increases. Notably, at a significant distance from the Sun (R\textsubscript{Sun}), KAWs exhibit a pronounced ability to transfer power throughout the loop. In the solar environment closer to the Sun, at h = 0.05 R\textsubscript{Sun} (left panel), KAWs effectively transport power over a larger distance. Conversely, at h = 0.1 R\textsubscript{Sun} (right panel), the power transfer occurs over a relatively shorter distance. This rapid energy transport suggests that in regions like the solar corona, KAWs efficiently deliver substantial energy across distances that extend beyond the flux tube's confines. The shifts observed in the curves for different $\mathrm{T_e/T_i}$ ratios provide insights into the wave-particle interactions within the flux tube loop. This phenomenon is attributed to the distinctive characteristics of KAWs' power delivery rate, which enhances wave-particle resonance and energy transfer, thereby facilitating more effective energy transport processes.

\subsection{Analysis of IAWs}
\begin{figure}
        \centering
       {
            \includegraphics[width=.47\linewidth]{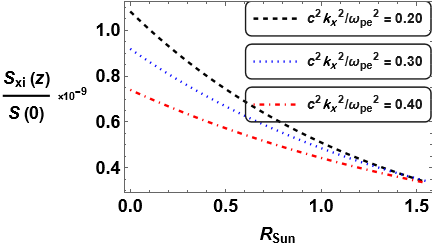}
                    }\quad
        {
            \includegraphics[width=.47\linewidth]{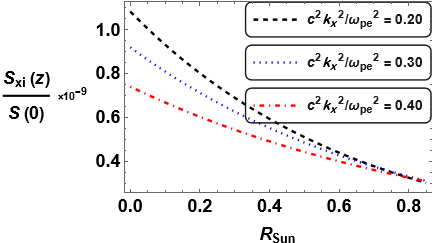}
            }
        \caption{Normalized perpendicular Poynting flux vector $S_{xi} (z)/S(0)$ of IAWs as a function of distance in the solar radii R\textsubscript{Sun}, evaluated for various inertial length values $c k_x/\omega_{pe}$. These plots, based on Eqn. (\ref{eq:23}) and the parameters are the same as in Figure. (\ref{B}), illustrate the impact of different $c k_x/\omega_{pe}$ values on the flux rate. In the left panel, h = 0.05 R\textsubscript{Sun} and h = 0.01 R\textsubscript{Sun} in the right panel.}
    \label{G}
\end{figure}

In our recent work \citep{batool2024acceleration}, we evaluated the parallel Poynting flux of IAWs across various parameters. However, the perpendicular flux of IAWs remains unexplored in existing literature. In this study, we addressed this gap by analyzing the perpendicular Poynting flux for different values of the normalized inertial length $c k_x/ \omega_{pe}$, as illustrated in Figure \ref{G}. We find that the magnitude of the perpendicular Poynting flux decreases with increasing $c k_x/\omega_{pe}$ values. This indicates that the perpendicular flux dissipates more rapidly, allowing IAWs to heat the plasma over shorter distances.

Notably, in regions with smaller heights, such as h = 0.05 R\textsubscript{Sun} (Figure \ref{G} left panel), IAWs can transport energy over a relatively large distance before being fully dissipated. In contrast, at greater heights of h = 0.1 R\textsubscript{Sun} (right panel), IAWs lose their energy much quicker, limiting their ability to heat the plasma over long distances. This highlights the efficiency of IAWs in delivering energy in smaller coronal regions close to the Sun's surface, where they play a crucial role in plasma heating.

\begin{figure}
    \centering
    \includegraphics[width=0.5\linewidth]{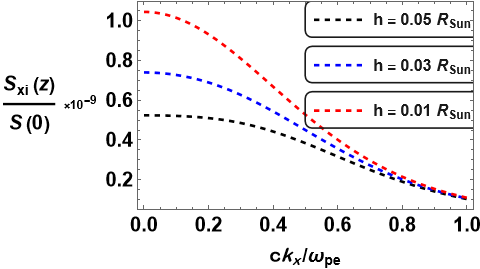}
    \caption{Normalized $S_{xi} (z)/S(0)$ of IAWs as a function of the normalized inertial length $c k_x/\omega_{pe}$ for different values of h.}
    \label{H}
\end{figure}

We also analyzed the perpendicular heating effects of IAWs across varying heights (h). The results show that the magnitude of $S_{xi} (z)/S(0)$ increases with greater heights (see Figure \ref{H}). Specifically, in regions with longer heights, such as h = 0.25 R\textsubscript{Sun}, the Poynting flux associated with IAWs dissipates more slowly compared to shorter heights (i.e., h = 0.05 R\textsubscript{Sun}). In contrast, the flux rate decays much faster in shorter-height regions, indicating that IAWs dissipate more efficiently in these areas. This rapid dissipation in smaller coronal regions leads to more intense plasma heating, highlighting the significant role of IAWs in heating the solar corona, especially in areas closer to the Sun's surface.

\begin{figure}
    \centering
    \includegraphics[width=0.5\linewidth]{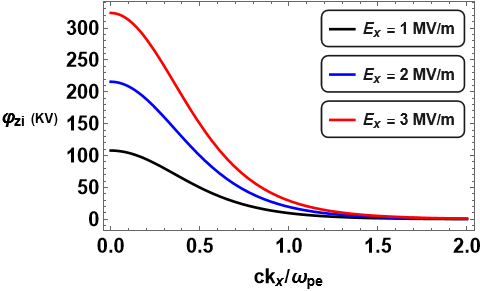}
    \caption{The parallel electric potential $\phi_{zi}$ of IAWs as a function of the normalized inertial length $c k_x/\omega_{pe}$ for different values of the electric field $\mathrm{E_x}$. The plots are based on Eq. (\ref{eq:1c}) with the same parameters as we used in Figure \ref{B} except $k_x \approx \times 10^{-3}$ cm\textsuperscript{-1}.}
    \label{Ha}
\end{figure}

The electric potential $\mathrm{\phi_{iz}}$ associated with IAWs is shown in Figure \ref{Ha}. The figure reveals a clear trend: the amplitude of $\phi_{iz}$ significantly increases with even a small increment in the perpendicular electric field $\mathrm{E_x}$. This suggests that as $\mathrm{E_x}$ increases, the energy associated with the potential in the parallel direction rises, enhancing the ability of IAWs to interact with charged particles. However, it is important to note that as the normalized perpendicular electron inertial length $c k_x/\omega_{pe}$ increases beyond a certain point, the magnitude of $\mathrm{\phi_{iz}}$ approaches zero. This implies that the influence of IAWs, in terms of electric potential, is confined to a limited range of $c k_x/\omega_{pe}$ values. Outside this range, the IAWs' ability to maintain a significant potential diminishes, effectively restricting the wave's influence to shorter scales.

Despite this limitation, the electric potential of IAWs in the solar corona can be quite large and carry a substantial amount of energy. This energy can be crucial in accelerating electrons along the magnetic field lines, as highlighted by \cite{watt2006inertial}. In this context, IAWs become important contributors to particle dynamics, particularly for electron acceleration in field-aligned directions. In the inertial regime, the parallel electric field component ($E_z$) is smaller compared to the perpendicular component ($\mathrm{E_x}$), as noted by \cite{mcclements2009inertial}. This discrepancy suggests that, although the energy transported by IAWs is mainly concentrated in the perpendicular electric field, the small parallel electric field still plays a critical role in accelerating particles along the magnetic field.

Further reinforcing this, \cite{fletcher2008impulsive} observed that the perpendicular electric field of an Alfv\'{e}n wave in the solar corona can reach extremely high values, up to an order of $10^6$ Vm\textsuperscript{-1}, whereas, the parallel electric fields are much smaller—typically less than 1 Vm\textsuperscript{-1} — they are still sufficient to accelerate electrons to relativistic energies in the field-aligned direction. The existence of such small, yet effective, parallel electric fields shows the powerful role of IAWs in particle energization, particularly in the corona where the magnetic fields are strong and the particle populations are dynamic.

Supporting this view, \cite{stasiewicz2008electric} studied electric potentials and energy fluxes related to particle acceleration in the solar corona and found that slow Alfv\'{e}nons generate electric potentials on the order of several kilovolts (kV). These potentials can accelerate solar wind ions to high energies, further emphasizing the critical role of wave-particle interactions in plasma acceleration. More recently, \cite{kostarev2021alfven} investigated the parallel electric fields of Alfv\'{e}n waves in a dipole model of the magnetosphere. They found that the parallel electric field potential associated with typical Alfv\'{e}n wave amplitudes observed in space can reach several kV, providing sufficient acceleration for auroral electrons. This demonstrates that even modest parallel electric fields can profoundly affect charged particle dynamics.

In our analysis, we used these established values of electric fields and the electric potentials to compare with our analytical results. Remarkably, our results align well with these observations, confirming the consistency of our model with real-world data. This consistency underscores the accuracy and reliability of our approach in capturing the dynamics of IAWs and their role in particle acceleration within the solar corona.

\begin{figure}
    \centering
    \includegraphics[width=0.5\linewidth]{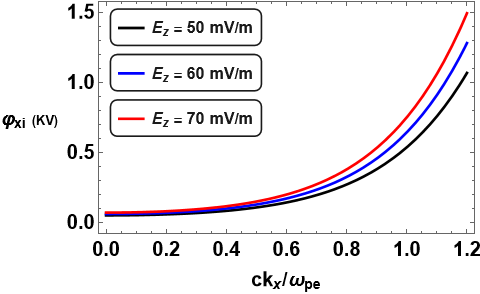}
    \caption{The perpendicular electric potential $\mathrm{\phi_{xi}}$ of IAWs as a function of the normalized inertial length $c k_x/\omega_{pe}$ for different values of the electric field $\mathrm{E_x}$. The plots are based on Eqn. (\ref{eq:1d}) with the same parameters as we used in Figure \ref{Ha}.}
    \label{Hb}
\end{figure}

We also evaluate the perpendicular electric potential $\mathrm{\phi_{xi}}$ of IAWs for different values of the parallel electric field $\mathrm{E_z}$, as illustrated in Figure \ref{Hb}. The results indicate that the magnitude of $\mathrm{\phi_{xi}}$ increases substantially with an increase in either $\mathrm{E_z}$ or the normalized $c k_x/\omega_{pe}$. This implies that the perpendicular potential becomes more pronounced in the larger $c k_x/\omega_{pe}$ regimes. A higher $\mathrm{E_z}$ corresponds to more energy being injected into the wave in the parallel direction, which, in turn, influences the perpendicular electric potential. A stronger parallel electric field directly contributes to a higher perpendicular potential, suggesting a strong coupling between these components in the inertial Alfv\'{e}n wave regime.

These findings suggest that IAWs can effectively generate substantial perpendicular electric potentials. This enhanced potential in turn affects the dynamics of particles in the plasma, particularly in terms of perpendicular heating and acceleration.

\begin{figure}
        \centering
       {
            \includegraphics[width=.47\linewidth]{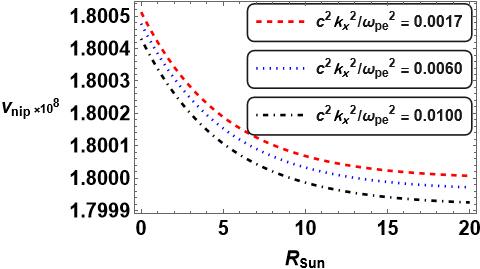}
                    }\quad
        {
            \includegraphics[width=.47\linewidth]{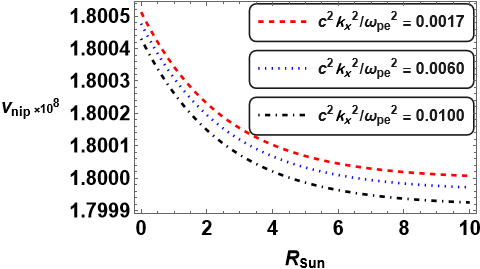}
            }
        \caption{The perpendicular net speed $\mathrm{V_{nip}}$ of particles versus distance R\textsubscript{Sun} for different values of the inertial length $c k_x/\omega_{pe}$. The plots are based on Eqn. (\ref{eq:23}) with the same parameters we used in Figure \ref{B}. except $k_x \sim \times 10^{-2}$ cm\textsuperscript{-1} and $k_z \sim \times 10^{-6}$ cm\textsuperscript{-1}. In the left panel, h = 0.05 R\textsubscript{Sun} and h = 0.01 R\textsubscript{Sun} in the right panel. The magnitude of $V_{nip}$ is weakly enhanced for different values of $c k_x/\omega_{pe}$.}
    \label{I}
\end{figure}

Figure \ref{I} illustrates the resultant net perpendicular speed $\mathrm{V_{nip}}$ of particles after wave-particle interaction for varying values of $c k_x/\omega_{pe}$. As $c k_x/\omega_{pe}$ increases, the magnitude of $\mathrm{V_{nip}}$ is slightly decreases. At a height of h = 0.05 R\textsubscript{Sun} (left panel), particles accelerate and heat the corona over a greater distance R\textsubscript{Sun}. Conversely, at h = 0.1 R\textsubscript{Sun} (right panel), heating occurs over a shorter range, with $V_{nip}$ decaying more rapidly in smaller R\textsubscript{Sun} regions. IAWs interact with particles and dissipate, transferring energy at resonance, which leads to significant particle acceleration and extended coronal heating.

Moreover, we also examined $\mathrm{V_{nip}}$ across different h and $\mathrm{T_e/T_i}$ values. We find no significant variations in the net perpendicular speed of IAWs. Explicitly, the height and $\mathrm{T_e/T_i}$ can influence $\mathrm{V_{nip}}$, their effects are subtle. For instance, the expression (\ref{eq:26}) shows a dependency on electron thermal velocity, where the temperature could alter $\mathrm{V_{nip}}$. However, this requires more sophisticated analysis, i.e., rewriting the thermal velocity in terms of temperature which we discussed in our recent research work, for instance, see \citep{batool2024acceleration}. Regarding height, since we are investigating IAWs in the solar corona, focusing on smaller h regions near the Sun, thus smaller h values do not significantly affect the net perpendicular speed of IAWs. However, at far proximity of the Sun's environment, where h values are larger, $\mathrm{V_{nip}}$ is influenced by larger h values which are beyond the scope of our chosen solar coronal regimes.

\begin{figure}
        \centering
       {
            \includegraphics[width=.47\linewidth]{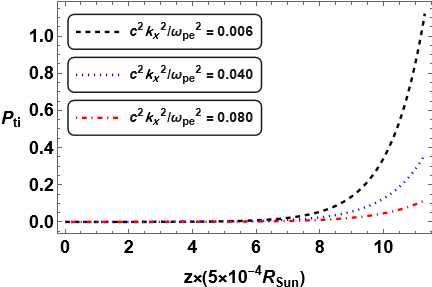}
                    }\quad
        {
            \includegraphics[width=.47\linewidth]{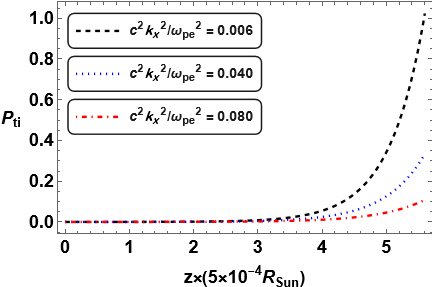}
            }
        \caption{The power transfer rate of IAWs as a function of the normalized distance $z\times (0.0005$ R\textsubscript{Sun}) for different $c k_x/\omega_{pe}$ values. The parameter values are the ones we used in the Figure \ref{I}. In the left panel, h = 0.05 R\textsubscript{Sun} and h = 0.01 R\textsubscript{Sun} in the right panel.}
    \label{J}
\end{figure}

The power transfer rate of IAWs in the solar flux loop tube is illustrated in Figure \ref{J}. As we examine the effects of the normalized inertial length $c k_x/\omega_{pe}$, it becomes evident that the power rate of IAWs decreases significantly. This reduction is most pronounced at larger distances from the Sun (R\textsubscript{Sun}), where the curve differences are more visible. In the environment close to the Sun, specifically at h = 0.05  R\textsubscript{Sun} (left panel), the net power of IAWs through the solar flux tube initiates at a greater distance, approximately after 6  R\textsubscript{Sun}. Conversely, when h = 01  R\textsubscript{Sun} (right panel), the onset of this power shifting occurs at nearly half the distance compared to the former scenario. This indicates that in regions closer to the Sun, IAWs are more effective in transporting power over larger distances within the solar coronal flux loop tube. However, as we move farther from the Sun, at h = 0.1  R\textsubscript{Sun}, the power transfer by IAWs begins at shorter distances.

Furthermore, the magnitude and shifting in the power rate of IAWs diminish for larger values of $c k_x/\omega_{pe}$, as seen in Figure \ref{J} (i.e., see the black curves). These observations suggest that as the distance from the Sun increases, the efficiency of IAWs in transferring power diminishes, particularly at greater inertial lengths. This phenomenon underscores the varying dynamics of wave-particle interactions within the solar coronal flux loop tube, where IAWs play a critical role in modulating energy transfer processes over different solar distances.

\section{Summary}
The study highlights the intricate dynamics of KAWs and IAWs within the solar corona and solar wind, focusing on their role in heating and accelerating charged particles. A significant aspect of our research is the emphasis on parallel and perpendicular electric potential associated with KAWs and IAWs, perpendicular heating, power deposition, and particle acceleration. Our findings indicate that the parallel Poynting flux of KAWs dissipates more rapidly with increasing $\mathrm{T_e/T_i}$. In contrast, the perpendicular Poynting flux strengthens and becomes less prone to dissipation at larger solar distances R\textsubscript{Sun}. This suggests that as $\mathrm{T_e/T_i}$ rises, KAWs contribute more effectively to perpendicular heating, particularly in regions far from the Sun. Notably, for higher values of $k_x \rho_i$, the perpendicular Poynting flux of KAWs exhibits a more rapid dissipation at elevated $\mathrm{T_e/T_i}$ ratios, showcasing the sensitivity of these waves to temperature variations.

In the case of IAWs, our analysis reveals a decrease in perpendicular Poynting flux as the inertial length $c k_x/\omega_{pe}$ increases, suggesting that IAWs dissipate their energy quickly in the perpendicular direction, with their impact being most significant in the near-Sun (R\textsubscript{Sun}) regions. A higher value of $c k_x/\omega_{pe}$ corresponds to shorter perpendicular wavelengths, which are more susceptible to localized damping, thus leading to reduced energy transport and weaker heating effects over extended distances. Conversely, IAWs with longer perpendicular wavelengths (smaller $c k_x/\omega_{pe}$) dissipate more gradually, allowing them to propagate further and contribute to coronal heating over a wider spatial domain. From a physical perspective, the mathematical expression (Eqn. \ref{eq:23})  for the perpendicular Poynting flux reveals an explicit dependence on electron temperature through the electron thermal velocity term. This highlights the role of thermal anisotropy, as the temperature of plasma species (particularly electrons) is not typically uniform along and across the magnetic field lines. Specifically for IAWs, where the electron temperature $T_e$ exceeds the ion temperature $T_i$, the higher perpendicular thermal energy associated with electrons intensifies damping across the magnetic field. This leads to efficient yet localized heating of the solar corona over short radial distances. IAWs are highly effective in transferring and dissipating energy perpendicularly, especially in regions closer to the Sun, contributing to localized coronal heating.

The parallel and perpendicular electric potentials associated with both KAWs and IAWs are evaluated for various values of the parallel and perpendicular electric fields ($\mathrm{E_z}$ and $\mathrm{E_x}$). Our analysis reveals that the electric potentials generated by these waves are sufficiently strong to accelerate particles in the solar corona. In particular, the perpendicular potentials, though smaller in magnitude than their parallel counterparts, are crucial for driving particle acceleration across the magnetic field lines. These combined effects of electric potentials indicate that both KAWs and IAWs play a vital role in energizing particles, leading to enhanced heating and acceleration in the solar coronal environment.

Beyond the Poynting flux vectors and the electric potentials, we also explored the net resonant speed of particles. For KAWs, we found that the perpendicular resonance speed diminishes with higher $\mathrm{T_e/T_i}$ ratios, enabling particles to accelerate and heat the corona over more extended distances from the Sun. Conversely, for IAWs, the magnitude of the perpendicular net resonance speed decreases with increasing inertial length. However, we observed that the particles' acceleration because of the IAWs can heat up the solar corona over more extended distances. Thus, IAWs are more effective for local heating near the base of the solar corona, where the inertial length is smaller and wave-particle coupling is more effective.

Finally, we investigated the total power transfer rate of both KAWs and IAWs within the solar flux loop tube. Our analysis reveals that the net power transfer by KAWs increases with higher values of the electron-to-ion temperature ratio ($\mathrm{T_e/T_i}$). KAWs exhibit the ability to transport significant power across the flux loop, particularly at larger distances from the Sun, denoted by R\textsubscript{Sun}. This suggests that KAWs are highly effective in delivering energy over extended distances within the solar corona, especially in regions where the temperature ratio is elevated. Conversely, the power delivery by IAWs is found to diminish as the normalized inertial length parameter $c k_x/ \omega_{pe}$ increases. Notably, IAWs display a more pronounced power transfer rate at greater R\textsubscript{Sun} distances, indicating that these waves play a more substantial role in energy transport within the outer regions of the solar flux loop tube. The contrasting behaviors of KAWs and IAWs concerning their respective parameters underscore the complex dynamics of wave-particle interactions and energy transfer processes in the solar corona.

Our research offers valuable insights into the behavior of KAWs and IAWs in the solar corona and solar wind, with potential applications extending to other space plasma environments. Future work could enhance this model by considering temperature anisotropy and incorporating non-Maxwellian distribution functions, providing a more comprehensive understanding of wave-particle interactions in space plasmas.

\subsection{Acknowledgments}
SA acknowledges the support of an NSF grant 2149771 and GPZ the partial support of a NASA Parker Solar Probe contract SV4 - 84017 and an NSF EPSCoR RII - Track - 1 Cooperative Agreement OIA - 2148653.

%%%%%%%%%%%%%%%%%%%%%%%%%%%%%%%%%%%%%%%%%%%%%%%%%%
\section*{Data Availability}
All data generated and/or analyzed during this study are included in the article. The work consists of detailed theoretical derivations, which will be made available (in step-wise form) upon reasonable request to the corresponding author, S. Ayaz.




%%%%%%%%%%%%%%%%%%%%%%%%%%%%%%%%%%%%%%%%%%%%%%%%%%

%%%%%%%%%%%%%%%%% APPENDICES %%%%%%%%%%%%%%%%%%%%%

%\appendix



%%%%%%%%%%%%%%%%%%%%%%%%%%%%%%%%%%%%%%%%%%%%%%%%%%

% Don't change these lines
\bsp	% typesetting comment
\label{lastpage}

\begin{thebibliography}{}
\providecommand{\natexlab}[1]{#1}
\providecommand{\url}[1]{\texttt{#1}}
\expandafter\ifx\csname urlstyle\endcsname\relax
  \providecommand{\doi}[1]{doi: #1}\else
  \providecommand{\doi}{doi: \begingroup \urlstyle{rm}\Url}\fi

\end{thebibliography}

\begin{thebibliography}{}

\bibitem[Van der Holst et al. (2014)]{van2014alfven}
Van der Holst, B., Sokolov, I. V., Meng, X., Jin, M., Manchester IV, W. B., Tóth, G., \& Gombosi, T. I. (2014). Alfv\'{e}n wave solar model (AWSoM): coronal heating. \textit{The Astrophysical Journal}, 782(2), 81.

\bibitem[De Pontieu et al.(2007)]{de2007chromospheric}
De Pontieu, B., McIntosh, S. W., Carlsson, M., Hansteen, V. H., Tarbell, T. D., Schrijver, C. J., Title, A. M., Shine, R. A., Tsuneta, S., Katsukawa, Y., et al. (2007). Chromospheric Alfv\'{e}nic waves strong enough to power the solar wind. \textit{Science}, 318(5856), 1574--1577.

\bibitem[McIntosh et al.(2011)]{mcintosh2011alfvenic}
McIntosh, S. W., De Pontieu, B., Carlsson, M., Hansteen, V., Boerner, P., \& Goossens, M. (2011). Alfv\'{e}nic waves with sufficient energy to power the quiet solar corona and fast solar wind. \textit{Nature}, 475(7357), 477--480.

\bibitem[Sokolov et al.(2013)]{sokolov2013magnetohydrodynamic}
Sokolov, I. V., Van der Holst, B., Oran, R., Downs, C., Roussev, I. I., Jin, M., Manchester, W. B., Evans, R. M., \& Gombosi, T. I. (2013). Magnetohydrodynamic waves and coronal heating: unifying empirical and MHD turbulence models. \textit{The Astrophysical Journal}, 764(1), 23.

\bibitem[Coleman et al. (1968)]{coleman1968turbulence}
Coleman Jr, P. J. (1968). Turbulence, viscosity, and dissipation in the solar-wind plasma. \textit{The Astrophysical Journal}, 153, 371.

\bibitem[Belcher et al.(1971)]{belcher1971large}
Belcher, J. W., \& Davis Jr, L. (1971). Large-amplitude Alfv\'{e}n waves in the interplanetary medium, 2. \textit{Journal of Geophysical Research}, 76(16), 3534--3563.

\bibitem[Alazraki et al.(1971)]{alazraki1971solar}
Alazraki, G., \& Couturier, P. (1971). Solar wind acceleration caused by the gradient of Alfv\'{e}n wave pressure. \textit{Astronomy and Astrophysics}, 13, 380.

\bibitem[Usmanov et al.(2000)]{usmanov2000global}
Usmanov, A. V., Goldstein, M. L., Besser, B. P., \& Fritzer, J. M. (2000). A global MHD solar wind model with WKB Alfv\'{e}n waves: Comparison with Ulysses data. *Journal of Geophysical Research: Space Physics*, 105(A6), 12675--12695.

\bibitem[Hu et al.(2003)]{hu2003coronal}
Hu, Y. Q., Habbal, S. R., Chen, Y., \& Li, X. (2003). Are coronal holes the only source of fast solar wind at solar minimum? \textit{ournal of Geophysical Research: Space Physics}, 108(A10).

\bibitem[Suzuki et al.(2006)]{suzuki2006solar}
Suzuki, T. K., \& Inutsuka, S. (2006). Solar winds driven by nonlinear low-frequency Alfv\'{e}n waves from the photosphere: Parametric study for fast/slow winds and disappearance of solar winds. \textit{Journal of Geophysical Research: Space Physics}, 111(A6).

\bibitem[Matsumoto(2012)]{matsumoto2012connecting}
Matsumoto, T., \& Suzuki, T. K. (2012). Connecting the Sun and the solar wind: The first 2.5-dimensional self-consistent MHD simulation under the Alfv\'{e}n wave scenario. \textit{The Astrophysical Journal}, 749(1), 8.

\bibitem[Heinemann et al.(1980)]{heinemann1980non}
Heinemann, M., \& Olbert, S. (1980). Non-WKB Alfv\'{e}n waves in the solar wind. \textit{Journal of Geophysical Research: Space Physics}, 85(A3), 1311--1327.

\bibitem[Leroy et al.(1980)]{leroy1980propagation}
Leroy, B. (1980). Propagation of waves in an atmosphere in the presence of a magnetic field. II—The reflection of Alfvén waves. \textit{Astronomy and Astrophysics}, 91(1–2), 136--146.

\bibitem[Matthaeus et al.(1999)]{matthaeus1999coronal}
Matthaeus, W. H., Zank, G. P., Oughton, S., Mullan, D. J., \& Dmitruk, P. (1999). Coronal heating by magnetohydrodynamic turbulence driven by reflected low-frequency waves. \textit{The Astrophysical Journal}, 523(1), L93.

\bibitem[Zank et al. (2018)]{zank2018theory}
Zank, G. P., Adhikari, L., Hunana, P., Tiwari, S. K., Moore, R., Shiota, D., Bruno, R., \& Telloni, D. (2018). Theory and transport of nearly incompressible magnetohydrodynamic turbulence. IV. Solar coronal turbulence. \textit{The Astrophysical Journal}, 854(1), 32.

\bibitem[Zank et al. (2021)]{zank2021turbulence}
Zank, G. P., Zhao, L.-L., Adhikari, L., Telloni, D., Kasper, J. C., \& Bale, S. D. (2021). Turbulence transport in the solar corona: Theory, modeling, and Parker Solar Probe. \textit{Physics of Plasmas}, 28(8).

\bibitem[Velli et al.(1989t)]{velli1989turbulent}
Velli, M., Grappin, R., \& Mangeney, A. (1989). Turbulent cascade of incompressible unidirectional Alfv\'{e}n waves in the interplanetary medium. \textit{Physical Review Letters}, 63(17), 1807.

\bibitem[Parker et al.(1958)]{parker1958dynamics}
Parker, E. N. (1958). Dynamics of the interplanetary gas and magnetic fields. \textit{Astrophysical Journal}, 128, 664.

\bibitem[Hundhausen et al.(1968)]{hundhausen1968direct}
Hundhausen, A. J. (1968). Direct observations of solar-wind particles. \textit{Space Science Reviews}, 8(5), 690--749.

\bibitem[Parker et al.(2019)]{parker2019cosmical}
Parker, E. N. (2019). \textit{Cosmical magnetic fields: Their origin and their activity}. Oxford University Press.

\bibitem[Okamoto et al.(2007)]{okamoto2007coronal}
Okamoto, T. J., Tsuneta, S., Berger, T. E., Ichimoto, K., Katsukawa, Y., Lites, B. W., Nagata, S., Shibata, K., Shimizu, T., Shine, R. A., et al. (2007). Coronal transverse magnetohydrodynamic waves in a solar prominence. \textit{Science}, 318(5856), 1577--1580.

\bibitem[Cirtain et al.(2007)]{cirtain2007evidence}
Cirtain, J. W., Golub, L., Lundquist, L., Van Ballegooijen, A., Savcheva, A., Shimojo, M., DeLuca, E., Tsuneta, S., Sakao, T., Reeves, K., et al. (2007). Evidence for Alfv\'{e}n waves in solar X-ray jets. \textit{Science}, 318(5856), 1580--1582.

\bibitem[De et al.(1994)]{de1994coronal}
De Azevedo, C. A., Elfimov, A. G., \& De Assis, A. S. (1994). Coronal loop heating by Alfv\'{e}n waves. \textit{Solar Physics}, 153, 205--215.

\bibitem[Voitenko et al.(1995)]{voitenko1995anomalous}
Voitenko, Y. M. (1995). Anomalous magnetic diffusion in coronal current layers. \textit{Solar Physics}, 161, 197--200.

\bibitem[Voitenko et al.(1996)]{voitenko1996flare}
Voitenko, Y. M. (1996). Flare loops heating by the 0.1--1.0 MeV proton beams. \textit{Solar Physics}, 168, 219--222.

\bibitem[Elfimov et al.(1996)]{elfimov1996noninductive}
Elfimov, A. G., De Azevedo, C. A., \& De Assis, A. S. (1996). Noninductive current generation by Alfv\'{e}n wave-electron interaction in solar loops. \textit{Physica Scripta}, 1996(T63), 251.

\bibitem[Asgari et al.(2012)]{asgari2012model}
Asgari-Targhi, M., \& Van Ballegooijen, A. A. (2012). Model for Alfv\'{e}n wave turbulence in solar coronal loops: heating rate profiles and temperature fluctuations. \textit{The Astrophysical Journal}, 746(1), 81.

\bibitem[Morton et al.(2015)]{morton2015investigating}
Morton, R. J., Tomczyk, S., \& Pinto, R. (2015). Investigating Alfv\'{e}nic wave propagation in coronal open-field regions. *Nature Communications*, 6, 7813.

\bibitem[Testa et al.(2014)]{testa2014evidence}
Testa, P., De Pontieu, B., Allred, J., Carlsson, M., Reale, F., Daw, A., Hansteen, V., Martinez-Sykora, J., Liu, W., DeLuca, E. E., et al. (2014). Evidence of nonthermal particles in coronal loops heated impulsively by nanoflares. \textit{Science}, 346(6207), 1255724.

\bibitem[Singh et al.(2019)]{singh2019anisotropic}
Singh, H. D., \& Jatav, B. S. (2019). Anisotropic turbulence of kinetic Alfv\'{e}n waves and heating in solar corona. \textit{Research in Astronomy and Astrophysics}, 19(12), 185.

\bibitem[Cramer et al.(2011)]{cramer2011physics}
Cramer, N. F. (2011). \textit{The Physics of Alfv\'{e}n Waves}. John Wiley \& Sons.

\bibitem[Wu et al.(2020)]{wu2020kinetic}
Wu, D.-J., \& Chen, L. (2020). \textit{Kinetic Alfv\'{e}n Waves in Laboratory, Space, and Astrophysical Plasmas}. Springer.

\bibitem[Gekelman et al.(1999)]{gekelman1999review}
Gekelman, W. (1999). Review of laboratory experiments on Alfv\'{e}n waves and their relationship to space observations. \textit{Journal of Geophysical Research: Space Physics}, 104(A7), 14417--14435.

\bibitem[Stasiewicz et al.(2000)]{stasiewicz2000small}
Stasiewicz, K., Bellan, P., Chaston, C., Kletzing, C., Lysak, R., Maggs, J., Pokhotelov, O., Seyler, C., Shukla, P., Stenflo, L., et al. (2000). Small scale Alfv\'{e}nic structure in the aurora. \textit{Space Science Reviews}, 92, 423--533.

\bibitem[Wu et al.(2004)]{wu2004recent}
Wu, D. J., \& Chao, J. K. (2004). Recent progress in nonlinear kinetic Alfv\'{e}n waves. \textit{Nonlinear Processes in Geophysics}, 11(5/6), 631--645.

\bibitem[Keiling et al.(2009)]{keiling2009alfven}
Keiling, A. (2009). Alfv\'{e}n waves and their roles in the dynamics of the Earth’s magnetotail: a review. \textit{Space Science Reviews}, 142, 73--156.

\bibitem[Zhao et al.(2010)]{zhao2010nonlinear}
Zhao, J. S., Wu, D. J., \& Lu, J. Y. (2010). On nonlinear decay of kinetic Alfv\'{e}n waves and application to some processes in space plasmas. \textit{Journal of Geophysical Research: Space Physics}, 115(A12).

\bibitem[Wygant et al.(2000)]{wygant2000polar}
Wygant, J. R., Keiling, A., Cattell, C. A., Johnson, M., Lysak, R. L., Temerin, M., Mozer, F. S., Kletzing, C. A., Scudder, J. D., Peterson, W., et al. (2000). Polar spacecraft based comparisons of intense electric fields and Poynting flux near and within the plasma sheet-tail lobe boundary to UVI images: An energy source for the aurora. \textit{Journal of Geophysical Research: Space Physics}, 105(A8), 18675--18692.

\bibitem[Wygant et al.(2002)]{wygant2002evidence}
Wygant, J. R., Keiling, A., Cattell, C. A., Lysak, R. L., Temerin, M., Mozer, F. S., Kletzing, C. A., Scudder, J. D., Streltsov, V., Lotko, W., et al. (2002). Evidence for kinetic Alfv\'{e}n waves and parallel electron energization at 4--6 RE altitudes in the plasma sheet boundary layer. \textit{Journal of Geophysical Research: Space Physics}, 107(A8), SMP--24.

\bibitem[Lysak et al.(2003)]{lysak2003kinetic}
Lysak, R. L., \& Song, Y. (2003). Kinetic theory of the Alfv\'{e}n wave acceleration of auroral electrons. \textit{Journal of Geophysical Research: Space Physics}, 108(A4).

\bibitem[Gershman et al.(2017)]{gershman2017wave}
Gershman, D. J., Viñas, A. F., Dorelli, J. C., Boardsen, S. A., Avanov, L. A., Bellan, P. M., Schwartz, S. J., Lavraud, B., Coffey, V. N., Chandler, M. O., et al. (2017). Wave-particle energy exchange directly observed in a kinetic Alfv\'{e}n-branch wave. \textit{Nature Communications}, 8(1), 14719.

\bibitem[Tsiklauri et al.(2005)]{tsiklauri2005particle}
Tsiklauri, D., Sakai, J.-I., and Saito, S. (2005). Particle-In-Cell simulations of circularly polarised Alfv\'{e}n wave phase mixing: A new mechanism for electron acceleration in collisionless plasmas. \textit{Astronomy \& Astrophysics}, \textbf{435}(3), 1105--1113.

\bibitem[Genot et al.(1999)]{genot1999study}
Génot, V., Louarn, P., and Le Quéau, D. (1999). A study of the propagation of Alfvén waves in the auroral density cavities. \textit{Journal of Geophysical Research: Space Physics}, \textbf{104}(A10), 22649--22656.

\bibitem[Genot et al.(2004)]{genot2004alfven}
Génot, V., Louarn, P., and Mottez, F. (2004). Alfv\'{e}n wave interaction with inhomogeneous plasmas: acceleration and energy cascade towards small-scales. \textit{Annales Geophysicae}, \textbf{22}(6), 2081--2096.

\bibitem[Mottez et al.(2006)]{mottez2006comment}
Mottez, F., Génot, V., and Louarn, P. (2006). Comment on “PIC simulations of circularly polarised Alfv\'{e}n wave phase mixing...” by Tsiklauri et al. \textit{Astronomy \& Astrophysics}, \textbf{449}(2), 449--450.

\bibitem[Wwift et al.(2007)]{swift2007simulation}
Swift, D. W. (2007). Simulation of auroral electron acceleration by inertial Alfv\'{e}n waves. \textit{Journal of Geophysical Research: Space Physics}, \textbf{112}(A12).

\bibitem[Thompson et al.(1996)]{thompson1996electron}
Thompson, B. J., and Lysak, R. L. (1996). Electron acceleration by inertial Alfv\'{e}n waves. \textit{Journal of Geophysical Research: Space Physics}, \textbf{101}(A3), 5359--5369.

\bibitem[Damiano et al.(2023)]{damiano2023electron}
Damiano, P. A., Delamere, P. A., Kim, E.-H., Johnson, J. R., and Ng, C. S. (2023). Electron energization by inertial Alfv\'{e}n waves in density depleted flux tubes at Jupiter. \textit{Geophysical Research Letters}, \textbf{50}(5), e2022GL102467.

\bibitem[Szalay et al.(2020)]{szalay2020alfvenic}
Szalay, J. R., et al. (2020). Alfv\'{e}nic acceleration sustains Ganymede's footprint tail aurora. \textit{Geophysical Research Letters}, \textbf{47}(3), e2019GL086527.

\bibitem[Yalim et al.(2024)]{yalim2024mixing}
Yalim, J., et al. (2024). Mixing and transport of CO2 across a monolayer-covered surface in an open cylinder driven by a rotating knife edge. \textit{Physica D: Nonlinear Phenomena}, \textbf{463}, 134150.

\bibitem[Leer et al.(1982)]{leer1982acceleration}
Leer, E., Holzer, T. E., and Flå, T. (1982). Acceleration of the solar wind. \textit{Space Science Reviews}, \textbf{33}, 161--200.

\bibitem[Mottez et al.(2011)]{mottezelectron2011}
Mottez, F., and Génot, V. (2011). Electron acceleration by an Alfv\'{e}nic pulse. \textit{Journal of Geophysical Research}, \textbf{116}, A00K15.

\bibitem[Lysak et al.(1998)]{lysak1998relationship}
Lysak, R. L. (1998). The relationship between electrostatic shocks and kinetic Alfv\'{e}n waves. \textit{Geophysical Research Letters}, \textbf{25}(12), 2089--2092.

\bibitem[Lysak et al.(1996)]{lysak1996kinetic}
Lysak, R. L., and Lotko, W. (1996). On the kinetic dispersion relation for shear Alfv\'{e}n waves. \textit{Journal of Geophysical Research: Space Physics}, \textbf{101}(A3), 5085--5094.

\bibitem[Khan et al.(2020)]{khan2020solar}
Khan, I. A., Iqbal, Z., and Murtaza, G. (2020). Solar coronal heating by Alfv\'{e}n waves in bi-kappa distributed plasma. \textit{Monthly Notices of the Royal Astronomical Society}, \textbf{491}(2), 2403--2412.

\bibitem[Hasegawa et al.(1975)]{hasegawa1975kinetic}
Hasegawa, A., and Chen, L. (1975). Kinetic process of plasma heating due to Alfv\'{e}n wave excitation. \textit{Physical Review Letters}, \textbf{35}(6), 370.

\bibitem[Ayaz et al.(2024a)]{ayaz2024asolar}
Ayaz, S., Li, G., and Khan, I. A. (2024). Solar Coronal Heating by Kinetic Alfv\'{e}n Waves. \textit{The Astrophysical Journal}, \textbf{970}(2), 140.

\bibitem[Paraschiv et al.(2015)]{paraschiv2015physical}
Paraschiv, A. R., Bemporad, A., and Sterling, A. C. (2015). Physical properties of solar polar jets—A statistical study with Hinode XRT data. \textit{Astronomy \& Astrophysics}, \textbf{579}, A96.

\bibitem[Shukla et al.(2009)]{shukla2009study}
Shukla, N., Varma, P., and Tiwari, M. S. (2009). Study on kinetic Alfv\'{e}n wave in inertial regime. \textit{Indian Journal of Applied Physics}.

\bibitem[Lysak et al.(2023)]{lysak2023kinetic}
Lysak, R. L. (2023). Kinetic Alfv\'{e}n waves and auroral particle acceleration: A review. \textit{Reviews of Modern Plasma Physics}, \textbf{7}(1), 6.

\bibitem[Khan et al.(2019)]{khan2019distinct}
Khan, I. A., Khokhar, T. H., Shah, H. A., and Murtaza, G. (2019). Distinct features of Alfv\'{e}n wave in non-extensive plasmas. \textit{Physica A: Statistical Mechanics and its Applications}, \textbf{535}, 122385.

\bibitem[Khan et al.(2018)]{khan2018effect}
Khan, I. A., and Murtaza, G. (2018). Effect of kappa distribution on the damping rate of the obliquely propagating magnetosonic mode. \textit{Plasma Science and Technology}, \textbf{20}(3), 035302.

\bibitem[Li et al.(2023)]{li2023modeling}
Li, G., Shih, A. Y., Allen, R. C., Ho, G. C., Cohen, C. M. S., Desai, M., Dayeh, M. A., and Mason, G. M. (2023). Modeling Solar Energetic Neutral Atoms from Solar Flares and CME-driven Shocks. \textit{The Astrophysical Journal}, \textbf{944}(2), 196.

\bibitem[Batool et al.(2024)]{batool2024acceleration}
Batool, K., Khan, I. A., Shamir, M., Kabir, A., and Ayaz, S. (2024). Acceleration of solar wind particles due to inertial Alfv\'{e}n waves. \textit{Communications in Theoretical Physics}, \textbf{76}(6), 065501.

\bibitem[De Moortel et al.(2015)]{de2015recent}
De Moortel, I., and Browning, P. (2015). Recent advances in coronal heating. \textit{Philosophical Transactions of the Royal Society A}, \textbf{373}(2042), 20140269.

\bibitem[Zirin et al.(1996)]{zirin1996mystery}
Zirin, H. (1996). The mystery of the chromosphere. \textit{Solar Physics}, \textbf{169}, 313--326.

\bibitem[Gary et al.(2001)]{gary2001plasma}
Gary, G. A. (2001). Plasma beta above a solar active region: rethinking the paradigm. \textit{Solar Physics}, \textbf{203}, 71--86.

\bibitem[Pneuman et al.(1973)]{pneuman1973solar}
Pneuman, G. W. (1973). The solar wind and the temperature-density structure of the solar corona. \textit{Solar Physics}, \textbf{28}, 247--262.

\bibitem[Chen et al.(2012)]{chen2012kinetic}
Chen, L., and Wu, D. J. (2012). Kinetic Alfv\'{e}n wave instability driven by field-aligned currents in solar coronal loops. \textit{The Astrophysical Journal}, \textbf{754}(2), 123.

\bibitem[Goertz et al.(1984)]{goertz1984kinetic}
Goertz, C. K. (1984). Kinetic Alfv\'{e}n waves on auroral field lines. \textit{Planetary and Space Science}, \textbf{32}(11), 1387--1392.

\bibitem[Hasegawa et al.(1976)]{hasegawa1976kinetic}
Hasegawa, A., and Chen, L. (1976). Kinetic processes in plasma heating by resonant mode conversion of Alfv\'{e}n wave. Princeton Plasma Physics Lab. (Technical Report), Princeton, NJ (United States).

\bibitem[Goertz et al.(1979)]{goertz1979magnetosphere}
Goertz, C. K., and Boswell, R. W. (1979). Magnetosphere-ionosphere coupling. \textit{Journal of Geophysical Research: Space Physics}, \textbf{84}(A12), 7239--7246.

\bibitem[Lysak et al.(1993)]{lysak1993generalized}
Lysak, R. L. (1993). Generalized model of the ionospheric Alfv\'{e}n resonator. \textit{Geophysical Monograph Series}, \textbf{80}, 121--128.

\bibitem[Blanco et al.(2011)]{blanco2011electron}
Blanco-Benavides, J. M. (2011). Electron Acceleration by Inertial Alfv\'{e}n Waves.

\bibitem[Ayaz et al.(2020)]{ayaz2020alfven}
Ayaz, S., Khan, I. A., Iqbal, Z., and Murtaza, G. (2020). Alfv\'{e}n waves in temperature anisotropic Cairns distributed plasma. \textit{Communications in Theoretical Physics}, \textbf{72}(3), 035502.

\bibitem[Ayaz et al.(2019)]{ayaz2019dispersion}
Ayaz, S., Khan, I. A., and Murtaza, G. (2019). On the dispersion and damping of kinetic and inertial Alfv\'{e}n waves in Cairns distributed plasmas. \textit{Physics of Plasmas}, \textbf{26}(6).

\bibitem[Watt et al.(2005)]{watt2005self}
Watt, C. E. J., Rankin, R., Rae, I. J., and Wright, D. M. (2005). Self-consistent electron acceleration due to inertial Alfv\'{e}n wave pulses. \textit{Journal of Geophysical Research: Space Physics}, \textbf{110}(A10).

\bibitem[Chaston et al.(2000)]{chaston2000alfven}
Chaston, C. C., Carlson, C. W., Ergun, R. E., and McFadden, J. P. (2000). Alfv\'{e}n waves, density cavities and electron acceleration observed from the FAST spacecraft. \textit{Physica Scripta}, \textbf{2000}(T84), 64.

\bibitem[Fried et al.(2015)]{fried2015plasma}
Fried, B. D., and Conte, S. D. (2015). \textit{The Plasma Dispersion Function: The Hilbert Transform of the Gaussian}. Academic Press.

\bibitem[Richardson et al.(1995)]{richardson1995radial}
Richardson, J. D., Paularena, K. I., Lazarus, A. J., and Belcher, J. W. (1995). Radial evolution of the solar wind from IMP 8 to Voyager 2. \textit{Geophysical Research Letters}, \textbf{22}(4), 325--328.

\bibitem[Zhang et al.(2022)]{zhang2022observations}
Zhang, Z., Yuan, Z., Huang, S., Yu, X., Xue, Z., Deng, D., and Huang, Z. (2022). Observations of kinetic Alfv\'{e}n waves and associated electron acceleration in the plasma sheet boundary layer. \textit{Earth and Planetary Physics}, \textbf{6}(5), 465--473.

\bibitem[Kostarev et al.(2021)]{kostarev2021alfven}
Kostarev, D.~V., Mager, P.~N., \& Klimushkin, D.~Y. (2021). Alfv\'{e}n wave parallel electric field in the dipole model of the magnetosphere: gyrokinetic treatment. \textit{Journal of Geophysical Research: Space Physics}, \textit{126}(2), e2020JA028611.

\bibitem[Stasiewicz et al.(2008)]{stasiewicz2008electric}
Stasiewicz, K., \& Ekeberg, J. (2008). Electric potentials and energy fluxes available for particle acceleration by alfv\'{e}nons in the solar corona. \textit{The Astrophysical Journal}, \textit{680}(2), L153.

\bibitem[Watt et al.(2006)]{watt2006inertial}
Watt, C.~E.~J., Rankin, R., Rae, I.~J., \& Wright, D.~M. (2006). Inertial Alfv\'{e}n waves and acceleration of electrons in nonuniform magnetic fields. \textit{Geophysical Research Letters}, \textit{33}(2).

\bibitem[Mcclements et al.(2009)]{mcclements2009inertial}
McClements, K.~G., \& Fletcher, L. (2009). Inertial Alfv\'{e}n wave acceleration of solar flare electrons. \textit{The Astrophysical Journal}, \textit{693}(2), 1494.

\bibitem[Fletcher et al.(2008)]{fletcher2008impulsive}
Fletcher, L., \& Hudson, H.~S. (2008). Impulsive phase flare energy transport by large-scale Alfv\'{e}n waves and the electron acceleration problem. \textit{The Astrophysical Journal}, \textit{675}(2), 1645.

\bibitem[Alfv\'{e}n et al.(1942)]{alfven1942existence}
Alfvén, H. (1942). Existence of electromagnetic-hydrodynamic waves. \textit{Nature}, \textit{150}(3805), 405--406.

\bibitem[Shiota et al.(2017)]{shiota2017turbulent}
Shiota, D., Zank, G.~P., Adhikari, L., Hunana, P., Telloni, D., \& Bruno, R. (2017). Turbulent transport in a three-dimensional solar wind. \textit{The Astrophysical Journal}, \textit{837}(1), 75.

\bibitem[Zank et al.(2011)]{zank2011transport}
Zank, G.~P., Dosch, A., Hunana, P., Florinski, V., Matthaeus, W.~H., \& Webb, G.~M. (2011). The transport of low-frequency turbulence in astrophysical flows. I. Governing equations. \textit{The Astrophysical Journal}, \textit{745}(1), 35.

\bibitem[Osman et al.(2012)]{osman2012kinetic}
Osman, K.~T., Matthaeus, W.~H., Hnat, B., \& Chapman, S.~C. (2012). Kinetic signatures and intermittent turbulence in the solar wind plasma. \textit{Physical Review Letters}, \textit{108}(26), 261103.

\bibitem[Kiyani et al.(2015)]{kiyani2015dissipation}
Kiyani, K.~H., Osman, K.~T., \& Chapman, S.~C. (2015). Dissipation and heating in solar wind turbulence: from the macro to the micro and back again. \textit{Philosophical Transactions of the Royal Society A}, \textit{373}(2041), 20140155.

\bibitem[Ayaz et al.(2024b)]{ayaz2024balfven}
Ayaz, S., Zank, G.~P., Khan, I.~A., Li, G., \& Rivera, Y.~J. (2024). Alfv\'{e}nAlfv\'{e}n waves in the solar corona: resonance velocity, damping length, and charged particles acceleration by kinetic Alfv\'{e}n waves. \textit{Scientific Reports}, \textit{14}(1), 27275.

\bibitem[Ayaz et al.(2025)]{ayaz2025study}
Ayaz, S., Zank, G.~P., Khan, I.~A., Li, G., \& Rivera, Y.~J. (2025). A study of particle acceleration, heating, power deposition, and the damping length of kinetic Alfv\'{e}n waves in non-Maxwellian coronal plasma. \textit{Astronomy \& Astrophysics}, \textit{694}, A23.

\end{thebibliography}
\end{document}